\documentclass[11pt, a4paper]{article}
\pdfoutput=1
\usepackage{jcappub}

\usepackage{amssymb,amsmath,latexsym,mathrsfs}
\usepackage{graphicx} 
\usepackage{bm}
\usepackage{url}
\usepackage{epsfig,graphicx,verbatim,xspace,multirow,mathtools}
\usepackage{array}

\def\gsim{\mathrel{\raise.3ex\hbox{$>$\kern-.75em\lower1ex\hbox{$\sim$}}}}

\begin{document}



\hspace*{93mm}{\large \tt IFIC/18-09,  ULB-TH/18-02}\\

\title{A fresh look into the interacting dark matter scenario}

\author[a]{Miguel Escudero,}
\emailAdd{miguel.escudero@ific.uv.es}
\author[b]{Laura Lopez-Honorez,}
\emailAdd{llopezho@ulb.ac.be}
\author[a]{Olga Mena,}
\emailAdd{olga.mena@ific.uv.es}
\author[a]{Sergio Palomares-Ruiz,}
\emailAdd{sergio.palomares.ruiz@ific.uv.es}
\author[a]{Pablo Villanueva-Domingo}
\emailAdd{pablo.villanueva@ific.uv.es}

\affiliation[a]{Instituto de F\'isica Corpuscular (IFIC), CSIC-Universitat de Valencia,\\ 
Apartado de Correos 22085,  E-46071, Spain}

\affiliation[b]{Service de Physique Th\'eorique, CP225, Universit\'e Libre de Bruxelles, Bld du Triomphe, and Theoretische Natuurkunde, Vrije Universiteit Brussel and The International Solvay Institutes,\\
Pleinlaan 2, B-1050 Brussels, Belgium}

\abstract{The elastic scattering between dark matter particles and
  radiation represents an attractive possibility to solve a
  number of discrepancies between observations and standard cold dark
  matter predictions, as the induced collisional damping would imply a
  suppression of small-scale structures. We consider this scenario and
  confront it with measurements of the ionization history of the
  Universe at several redshifts and with recent estimates of the
  counts of Milky Way satellite galaxies. We derive a conservative
  upper bound on the dark matter-photon elastic scattering cross
  section of $\sigma_{\gamma \rm{DM}} < 8 \times 10^{-10} \, \sigma_T
  \, \left(m_{\rm DM}/{\rm GeV}\right)$ at $95\%$~CL, about one order
  of magnitude tighter than previous {constraints from satellite
    number counts}. Due to the strong degeneracies with astrophysical
  parameters, the bound on the dark matter-photon scattering cross
  section derived here is driven by the estimate of the number of
  Milky Way satellite galaxies. Finally, we also argue that future
  21~cm probes could help in disentangling among possible non-cold
  dark matter candidates, such as interacting and warm dark matter
  scenarios. Let us emphasize that bounds of similar magnitude to
    the ones obtained here could be also derived for models with dark
    matter-neutrino interactions and would be as constraining as the
    tightest limits on such scenarios.}
\maketitle

\section{Introduction}

The nature of the dark matter (DM) of our Universe still remains uncertain~\cite{Bertone:2004pz, Bergstrom:2009ib, Baer:2014eja, Bernal:2017kxu, Roszkowski:2017nbc, Buckley:2017ijx}. Despite the fact that the existence of non-relativistic, cold DM (CDM) is in excellent agreement with cosmic microwave background (CMB) and large scale structure (LSS) data~\cite{Ade:2015xua, Alam:2016hwk}, there are a number of observations at small scales, which may indicate a departure from this picture. This is commonly referred to as the "small-scale crisis" of the standard CDM paradigm~\cite{Bullock:2017xww}, which arises from the fact that CDM numerical predictions for the abundances and the kinematics of gravitationally bound structures show some discrepancies with observations~\cite{Klypin:1999uc, Moore:1999nt, BoylanKolchin:2011dk, Moore:1999gc, Springel:2008cc}. One of these discrepancies is the so-called ``missing satellite problem'': the fact that there are fewer observed satellites around our galaxy than the number predicted in CDM simulations~\cite{Klypin:1999uc, Moore:1999nt} (see however the recent Ref.~\cite{Kim:2017iwr} for a different perspective and reanalysis of this long-standing CDM problem). A number of dedicated simulations to set light on these issues have been carried out by different groups~\cite{Sawala:2012cn, Sawala:2015cdf, Fattahi:2016nld, Nakama:2017ohe}.

One very interesting possibility, extensively explored in the literature, is the case of the existence of DM candidates with non-negligible velocity dispersion at early times, which would result in the suppression of small-scale structures due to their free-streaming nature. Among this type of scenarios, warm DM (WDM) candidates with masses in the keV range could provide a solution to the missing satellite problem~\cite{Bode:2000gq, Knebe:2001kb, Colin:2007bk, Zavala:2009ms, Smith:2011ev, Lovell:2011rd, Schneider:2011yu, Polisensky:2013ppa, Lovell:2013ola, Kennedy:2013uta, Destri:2013hha, Angulo:2013sza, Benson:2012su, Kamada:2013sh, Lovell:2015psz, Ludlow:2016ifl, Wang:2016rio, Lovell:2016nkp, Nakama:2017ohe}. Indeed, the WDM hypothesis has received much attention due to claims of the detection of a monochromatic line at $(3.5-3.6)$~keV in X-ray data from galaxy clusters, the galactic center and the cosmic X-ray background~\cite{Bulbul:2014sua, Boyarsky:2014jta, Boyarsky:2014ska, Cappelluti:2017ywp}, which could potentially point to the radiative decay of a WDM keV sterile neutrino ($\nu_s \to \gamma \nu$). However, the most recent Lyman-$\alpha$ (Ly$\alpha$) forest constraints suggest $m_{\rm WDM} > (3-5)$~keV~\cite{Irsic:2017ixq, Yeche:2017upn} (for a thermal relic). In the case of sterile neutrino production via non-resonant active-sterile neutrino oscillations~\cite{Dodelson:1993je}, thermal equilibrium would never be reached, but one can relate the limit for thermal relics to the one for non-resonantly produced sterile neutrinos~\cite{Bozek:2015bdo}, resulting in $m_s \gtrsim (20-30)$~keV. 

Another very appealing scenario to solve the small-scale problem is to consider a DM candidate that interacts with relativistic particles (IDM for short) of the Standard Model (SM)~\cite{Boehm:2003xr,Boehm:2014vja, Schewtschenko:2014fca, Schewtschenko:2015rno} (i.e., photons or neutrinos, or from a hidden sector~\cite{Cyr-Racine:2013fsa, Buckley:2014hja, Vogelsberger:2015gpr, Cyr-Racine:2015ihg,Aarssen:2012fx,Bringmann:2016ilk}), with potential self-interactions~\cite{deLaix:1995vi, Spergel:1999mh, Kochanek:2000pi, Feng:2009mn, Loeb:2010gj, Vogelsberger:2012ku, Rocha:2012jg, Peter:2012jh, Dasgupta:2013zpn, Vogelsberger:2014pda, Cherry:2014xra, Vogelsberger:2015gpr, Lovell:2017eec} (see also Ref.~\cite{Murgia:2017lwo} for a generic treatment of the IDM and WDM effects under the name ``non-cold'' DM scenarios). Elastic scatterings between the DM particles and photons or neutrinos would give rise to collisional damping~\cite{Boehm:2000gq, Boehm:2001hm, Boehm:2004th}, which would erase small-scale structures and thus, could provide a mechanism to alleviate the aforementioned problems. 

In the following, we shall focus on a generic IDM model involving DM-photon interactions, characterized by the elastic scattering cross section $\sigma_{\gamma {\rm DM}}$. Even if the suppression of small-scale fluctuations in the IDM and WDM scenarios results from different physics (that is, collisional damping and free-streaming, respectively) it is possible to establish an approximate relation between the DM parameters that drive the small-scale suppression in these two scenarios (i.e.,  $\sigma_{\gamma {\rm DM}}$ and $m_{\rm WDM}$). Both the IDM and the WDM schemes have been exhaustively confronted against different cosmological  observations (CMB fluctuations and spectral distortions, galaxy clustering and Ly$\alpha$ forest power spectrum)~\cite{Boehm:2003xr, Boehm:2004th, Viel:2005qj, Seljak:2006qw, Viel:2006kd, Mangano:2006mp, Viel:2007mv, Boyarsky:2008xj, Serra:2009uu, McDermott:2010pa, Viel:2013apy, Wilkinson:2013kia,  Dolgov:2013una, Wilkinson:2014ksa, Boehm:2014vja, Schneider:2014rda, Schewtschenko:2014fca, Escudero:2015yka, Ali-Haimoud:2015pwa, Baur:2015jsy, Schewtschenko:2015rno, Diamanti:2017xfo, Irsic:2017ixq, Yeche:2017upn, Gariazzo:2017pzb, Murgia:2017lwo, Diacoumis:2017hff, DiValentino:2017oaw, Campo:2017nwh, Stadler:2018jin}.

Notice that the DM model featuring a velocity-independent DM-$\gamma$ scattering cross section considered in our work would correspond to millicharged DM. Indeed, the DM-$\gamma$ cross section can be expressed as
\begin{align}
\sigma_{\gamma {\rm DM}} = \epsilon^2 \sigma_T \left(\frac{m_e}{m_{\rm DM}}\right)^2 ~, 
\end{align}
where $\epsilon = |q|/e$ with $q$ the DM electric charge and $m_e$ the electron mass. Stringent limits on such a model can be obtained using CMB and Ly$\alpha$ data~\cite{Dvorkin:2013cea} (see also Ref.~\cite{Vinyoles:2015khy} for a compilation of limits),
\begin{align}
\epsilon < 1.8\times 10^{-6}\left(\frac{m_{\rm DM}}{\rm GeV}\right)^{1/2} ~,
\end{align}
which are valid for $m_{\rm DM} \gtrsim$~MeV.

In this work, we revisit the constraints on these non-CDM scenarios from the comparison of the predicted number of Milky Way (MW) satellite galaxies to the number estimated from actual observations. We consider the MW satellites recently discovered by the Dark Energy Survey (DES)~\cite{Bechtol:2015cbp, Drlica-Wagner:2015ufc}, in addition to the eleven classical objects plus those detected by the Sloan Digital Sky Survey (SDSS)~\cite{Ahn:2012fh, Koposov:2009ru}. The additional satellites, together with the assessment of the probability for a subhalo to form a galaxy~\cite{Kim:2017iwr}, change the expected bounds on the DM-photon elastic scattering cross section as compared to the earlier analysis of Ref.~\cite{Schewtschenko:2015rno}, which we discuss. Moreover, we also study the impact, within these scenarios, of the small-scale suppression on the ionization history of the Universe. Because of the reduction of power at small scales, reionization would be delayed with respect to the standard CDM case~\cite{Sitwell:2013fpa, Dayal:2014nva, Bose:2016hlz, Lopez-Honorez:2017csg}. Whereas many works in the literature have been devoted to test the impact of WDM scenarios on reionization-related observables~\cite{Barkana:2001gr, Yoshida:2003rm, Somerville:2003sh, Yue:2012na, Pacucci:2013jfa, Mesinger:2013nua, Schultz:2014eia, Dayal:2014nva, Lapi:2015zea, Bose:2016hlz, Bose:2016irl, Corasaniti:2016epp,Menci:2016eui, Lopez-Honorez:2017csg, Villanueva-Domingo:2017lae, Das:2017nub}, the modification of the ionization history of the Universe produced in IDM scenarios has only been recently addressed within a model with DM-hidden sector interactions~\cite{Lovell:2017eec}. Here, we consider measurements of the optical depth from the last scattering surface to reionization and of the ionization fraction at several redshifts. Given the current precision on these reionization observables and the (not yet fully understood) reionization efficiencies and thresholds, satellite galaxies counts turn out to provide the most powerful probe (among those used here) for testing IDM (as well as WDM) models. 

A similar analysis could be performed for a DM coupling to dark radiation\footnote{Notice that in Ref.~\cite{Vogelsberger:2015gpr}, an attempt to model the halo mass function of ETHOS simulations is studied making use of a sharp power law cut-off, similarly to WDM. The authors, however, point out that this approximation ceases to be valid at low halo masses, as ETHOS models have more power in the primordial power spectrum in comparison with WDM at those scales. Ref.~\cite{Moline:2016pbm} precisely goes beyond such an approximation, providing the fit~(\ref{eq:IDM}) for the DM-photon scattering case. Notice as well that the imprint of an ETHOS model on satellite number counts was considered in Ref.~\cite{Vogelsberger:2015gpr}, while its imprint on reionization observables was studied in Ref.~\cite{Lovell:2017eec}. } or neutrinos. Unfortunately, the information necessary to perform the analyses in this work is not publicly available.  Notice, though, that Ref.~\cite{Schewtschenko:2014fca} shows the strong (qualitative and quantitative) similarities between the DM-neutrino and DM-photon interaction models. We will use those results to provide a rough estimate of the bounds on DM-neutrino interactions that could be obtained using the same method as the one presented here.

The structure of the paper is as follows. In Section~\ref{sec:non-standard-dark}, we describe the halo mass function obtained from numerical simulations of WDM and IDM scenarios and discuss the relation between $\sigma_{\gamma {\rm DM}}$ and $m_{\rm WDM}$, based on similar small-scale suppression. This will be later used in the computation of the ionization history of the Universe in Section~\ref{sec:reio}, where we discuss how we calculate the evolution of the ionized fraction of the Universe and present the measurements we consider. In Section~\ref{sec:Nsat}, we discuss the main ingredients entering the computation of the number of MW satellites in WDM and IDM scenarios and indicate the current observational data. Finally, in Section~\ref{sec:results}, we show the constraints on the DM-photon cross section and on the WDM mass, using the different data sets individually and when combined. In that section, even if it is not the main focus of this paper, we also briefly discuss the prospects from future measurements of the 21~cm signal of neutral hydrogen and how these observations could be used to test and disentangle from each other non-CDM scenarios, as the ones studied in this work. Finally, in Section~\ref{sec:conclusions}, we summarize our results and draw our conclusions.

\section{Halo mass function of non-standard dark matter scenarios}
\label{sec:non-standard-dark}

When studying the number of satellites of the MW or the ionization history of the Universe, the density perturbations at small scales have long gone non-linear and one has to make use of the results from high resolution N-body simulations in order to account for the properties of DM halos. The halo mass function, that counts the number of halos per unit halo mass and volume at a given redshift, can be written as~\cite{Press:1973iz}
\begin{equation}
  \frac{dn}{dM}=-\frac12 \, \frac{\rho_m}{M^2} \, f(\nu) \, \frac{d\ln \sigma^2}{d\ln M} ~,
\label{eq:dndM}
\end{equation}
where $n$ is the halo number density, $\rho_m = \Omega_m \, \rho_c$ is the average matter density in the Universe at $z = 0$ and $\sigma^2 = \sigma^2(M,z)$ is the variance of density perturbations, which is a function of the halo mass $M$ and redshift. The first-crossing distribution, $f (\nu)$, is expected to be a universal function of $\nu \equiv \delta_c^2/\sigma^2(M,z)$, with $\delta_c = 1.686$, the linearly extrapolated density for collapse at $z = 0$. From the Press-Schechter formalism~\cite{Press:1973iz, Bond:1990iw}, but considering an ellipsoidal collapse model, the Sheth-Tormen (ST) first-crossing distribution was obtained as~\cite{Sheth:1999mn, Sheth:1999su, Sheth:2001dp}
\begin{eqnarray}
f(\nu) = A \, \sqrt{\frac{2 \, q \, \nu}{\pi}} \left(1 + (q \, \nu)^{-p}\right) \, e^{-q \, \nu/2} ~, 
\label{eq:ST}
\end{eqnarray}
with $p = 0.3$, $q = 0.707$ and $A = 0.3222$. For the standard CDM scenario, we consider this first-crossing distribution.\footnote{Note, however, that the values of the parameters in Eq.~(\ref{eq:ST}) are different from the default ones in the {\tt 21cmFAST} code (used throughout this work) and thus, have been correspondingly modified.}

As for the non-CDM scenarios we consider, we use the halo mass functions that match IDM and WDM simulations, as reported in Refs.~\cite{Moline:2016fdo, Schneider:2011yu}. The dependence in the particular DM cosmological model is partially encoded in the variance of density perturbations. The redshift dependence is driven by the linear growth function, $D(z)$ normalized to 1 at $z = 0$, so that the root-mean-square (rms) density fluctuation is $\sigma(M, z) = \sigma(M, z = 0) \, D(z)$. The variance at $z = 0$, $\sigma(M) \equiv \sigma(M, z = 0)$, at a given scale $R$ can be expressed as
\begin{equation}
\sigma_X^2(M(R)) = \int \frac{d^3k}{(2\pi)^3} \, P_X(k) \, W^2(kR) ~, 
\label{eq:sig2}
\end{equation}
where $P_X(k)$ is the linear power spectrum at $z = 0$ for a given $X = \{\textrm{CDM, IDM or  WDM}\}$ cosmology and $W$ is the Fourier transform of a filter function, which is defined as a top-hat (TH) function in real space,
\begin{equation}
W_{\rm TH}(kR) = \frac{3}{kR} \, \left(\sin(kR)- kR\cos(kR)\right) ~.
\label{eq:TH}
\end{equation}
In this case, the mass $M$ associated to the scale $R$ is given by $M = \frac{4\pi}{3} \, \rho_m \, R^3$.

In order to parameterize the small scale suppression of the power spectrum within a given model $X = \{\textrm{IDM or WDM}\}$ with respect to the CDM case, one can express the ratio of CDM and $X$ power spectra in terms of the transfer function $T_X$, defined as\footnote{Note that this is not to be confused with the usual transfer function used in $P_{\rm{CDM}}(k)$, which encodes the information on the evolution of density perturbations at different scales with respect to the primordial power spectrum, in the standard CDM scenario.}
\begin{equation}
P_{X}(k) = P_{\rm{CDM}}(k) \, T^2_{X}(k) ~,
\label{eq:pwdm}
\end{equation}
which can be parameterized in terms of a finite set of parameters and physical inputs, such as the WDM mass or the DM-photon elastic scattering cross section in IDM scenarios (see also Ref.~\cite{Murgia:2017lwo} for a recent generalization to a larger set of non-CDM models). Here we consider the fits obtained in Ref.~\cite{Bode:2000gq} for WDM and in Ref.~\cite{Boehm:2001hm} for IDM using the full perturbation evolution through a Boltzmann solver code. In those works the transfer function is expressed as
\begin{equation}
T_{X}(k) = \left(1+ (\alpha_{X} k)^{2\mu}\right)^{-5/\mu} ~,
\label{eq:twdm}
\end{equation}
where $\mu$ is a dimensionless exponent which is obtained to be $\mu = 1.2$ and $\alpha_X$ is the breaking scale, whose parameterization in terms of the model parameters will be specified in the next subsections. 

In order to describe the suppression in the linear regime, one can consider the half-mode mass $M_{\rm hm}$, defined as the mass scale for which $T_{X}/T_{\rm CDM} = 1/2$ (i.e., $P_{X}/P_{\rm CDM} = 1/4$). Using the general fit to the transfer function, Eq.~(\ref{eq:twdm}), the half-mode length $\lambda_{\rm hm}$  ($\lambda \equiv 2\pi/k$) and the half-mode mass $M_{\rm hm}$ can be easily derived as
\begin{eqnarray}
\lambda_{\rm hm} & = & 2\, \pi \, \alpha_X \left(2^{\mu/5} - 1 \right)^{-1/(2 \, \mu)} ~, \nonumber \\[2ex]
M_{\rm hm} & \equiv & \frac{4 \, \pi}{3} \, \rho_m \, \left( \frac{\lambda_{\rm hm}}{2}\right)^3 ~.
\label{eq:Mhm}
\end{eqnarray}
%

\subsection{Warm dark matter scenarios}
\label{sec:warm-dark-matter}

In the case of WDM scenarios, Refs.~\cite{Schneider:2011yu, Schewtschenko:2014fca} obtained the breaking scale, $\alpha_{\rm{WDM}}$, parameterized as~\cite{Bode:2000gq}
\begin{equation}
\alpha_{\rm{WDM}} = 0.048 \left(\frac{{\rm keV}}{m_{\rm{WDM}}}\right)^{1.15} \left(\frac{\Omega_{\rm{WDM}}} {0.4}\right)^{0.15}\left(\frac{h}{0.65}\right)^{1.3} \, {\rm Mpc}/h ~,
\label{eq:alphWDM}
\end{equation}
in terms of the WDM mass $m_{\rm WDM}$. 

In order to match the results of N-body simulations, the WDM halo mass function can be expressed as
\begin{equation}
\frac{dn^{\rm WDM}}{dM} = \left(1 + \frac{M_{\rm hm}}{b \, M}\right)^{a}  \frac{dn^{\rm ST, \, WDM}}{dM} ~,
\label{eq:WDM}
\end{equation}
where an additional mass-dependent correction to the standard ST formalism, governed by two parameters, $a$ and $b$, had to be introduced to reproduce the results of simulations~\cite{Schneider:2011yu}. The function $\frac{dn^{\rm ST, \, WDM}}{dM}$ refers to the halo mass function obtained with a ST first-crossing distribution, as defined in Eq.~(\ref{eq:ST}), and  a linear matter power spectrum corresponding to the WDM case. It was shown by Ref.~\cite{Schewtschenko:2014fca} that the agreement with WDM simulations was largely improved for $b = 0.5$.\footnote{Notice there was a typo in Ref.~\cite{Schewtschenko:2014fca}, as clarified in Ref.~\cite{Moline:2016fdo}.} Therefore, in the following, when considering the WDM halo mass functions, we shall refer to Eq.~(\ref{eq:WDM}) with $a = -0.6$ and $b = 0.5$.

\subsection{Interacting dark matter scenarios}

\begin{figure}[t]
\centering
\includegraphics[width=0.75\textwidth]{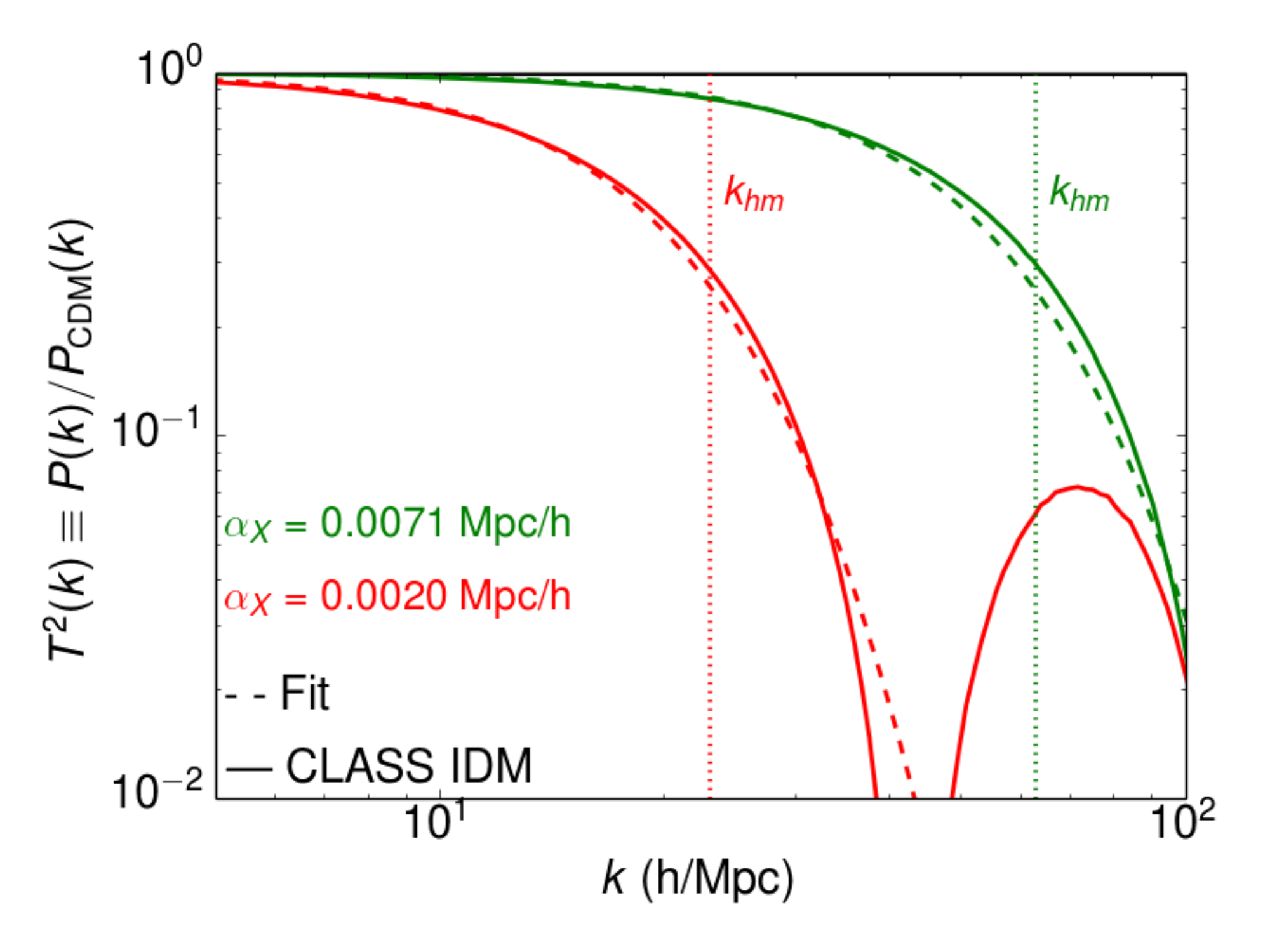} 
\caption{Transfer functions for the IDM scenario calculated with the Boltzmann solver CLASS~\cite{Lesgourgues:2011re} (solid curves) and for the WDM and IDM scenarios using the fitting formula in Eq.~(\ref{eq:twdm}), with Eqs.~(\ref{eq:alphWDM}) and (\ref{eq:alphIDM}) (dashed curves). We compare the results for two cases (see Tab.~\ref{tab:tabbench}) corresponding to the same half-mode length (i.e., the same half-mode mass $M_{\rm hm}$) or, equivalently, to the same breaking scale $\alpha_X = 0.0020$~Mpc/$h$ (leftmost red curves) and $\alpha_X = 0.0071$~Mpc/$h$ (rightmost green curves). The corresponding values for the DM-photon elastic scattering cross section in IDM and the DM mass in WDM scenarios are: $\sigma_{\gamma {\rm DM}} = 6.3 \times 10^{-10} \, \sigma_T \, (m_{\rm DM}/{\rm GeV})$ and $m_{\rm WDM} = 2.15$~keV (leftmost red curves) and $\sigma_{\gamma {\rm DM}} = 7.9 \times 10^{-11} \, \sigma_T \, (m_{\rm DM}/{\rm GeV})$ and $m_{\rm WDM} = 5.17$~keV (rightmost green curves). The wave-mode number corresponding to the half-mode length in each case is depicted by the dotted vertical lines. Notice that the agreement is very good until the typical damping oscillatory effects in IDM models start to dominate and create oscillations in the power spectrum.}
\label{fig:transfer}
\end{figure}

DM-photon interactions would give rise to large collisional damping effects that would also suppress the linear matter power spectrum at small scales~\cite{Boehm:2001hm, Boehm:2004th, Wilkinson:2014ksa, Wilkinson:2013kia, Schewtschenko:2015rno}. In Ref.~\cite{Boehm:2001hm}, such a reduction of the power at small scales was accounted for by means of Eq.~(\ref{eq:twdm}) with $\mu = 1.2$, and parameterizing the breaking scale in terms of the DM-photon elastic scattering cross section, 
\begin{equation}
\alpha_{\rm{IDM}}= 0.073 \, \left[10^8 \, \left(\frac{\sigma_{\gamma \rm{DM}}}{\sigma_T}\right) \, \left(\frac{{\rm GeV}}{m_{\rm DM}}\right)\right]^{0.48} \left(\frac{\Omega_{\rm{WDM}}} {0.4}\right)^{0.15}\left(\frac{h}{0.65}\right)^{1.3} \, {\rm Mpc}/h ~,
\label{eq:alphIDM}
\end{equation}
where $\sigma_T=6.65\times 10^{-25}$ cm$^2$ is the Thompson cross section. Notice that the suppression in the power spectrum at small scales can also be accounted for by DM-neutrino interactions, described with the same parameterization for the transfer function, with a small correction in the breaking scale, $\alpha_{\rm{\nu DM}}\simeq 0.8 \times \alpha_{\rm{\gamma DM}}$~\cite{Schewtschenko:2014fca}.

In Fig.~\ref{fig:transfer}, we show the results of the fitting formula for two benchmark models.  The curves for the transfer functions, obtained via Eq.~(\ref{eq:twdm}) (using Eqs.~(\ref{eq:alphWDM}) and~(\ref{eq:alphIDM}) for the WDM and IDM scenarios, respectively) are compared to the transfer function for the IDM case, computed numerically by means of the Boltzmann solver code CLASS~\cite{Lesgourgues:2011re}, where we have introduced the DM-photon interactions as described in Refs.~\cite{Wilkinson:2013kia, Boehm:2001hm}. Notice that this parameterization provides an accurate description of the IDM transfer function until the damped oscillatory effects at small scales appear. However, the accuracy of this description is enough for our purposes, as we are mostly interested in the region where the difference between IDM and CDM is maximal, and at the scale where the damped oscillations appear, the height of the second maximum is already suppressed by more than one order of magnitude (see Fig.~\ref{fig:transfer} and also, e.g., Ref.~\cite{Boehm:2001hm}).

Even if the fits for the power spectrum in WDM and IDM look very similar for masses above the half-mode mass, the oscillations in the power spectrum that appear at small scales in IDM scenarios introduce differences in the description of the number density of halos. As was noticed in Ref.~\cite{Schewtschenko:2014fca}, as a consequence of these oscillations, the number of low-mass structures in IDM is larger than in WDM scenarios and in order to reproduce the IDM results for masses below the half-mode mass, and extra mass-dependent correction must be introduced to the halo mass function~\cite{Moline:2016fdo},
\begin{equation}
\frac{dn^{\rm IDM}}{dM} =\left(1 + \frac{M_{\rm hm}}{b \, M}\right)^{a} \left(1+\frac{M_{\rm hm}}{g \, M}\right)^{c} \, \frac{dn^{\rm ST,\, CDM}}{dM} ~,
\label{eq:IDM}
\end{equation}
with $a = -1$, $b = 0.33$, $g = 1$, $c = 0.6$ and $\frac{dn^{\rm ST, \, CDM}}{dM}$ refers to the standard ST first-crossing distribution as defined in Eq.~(\ref{eq:ST}) and considering the CDM linear power spectrum for the variance of density perturbations. This approach has been shown to provide an excellent match to numerical simulations at $z = 0$ for $ \sigma_{\gamma {\rm DM}} = 2.0 \times 10^{-9} \, \sigma_T \, (m_{\rm DM}/{\rm GeV})$.  Notice, though, that a wider range of DM-photon scattering cross sections were considered in Ref.~\cite{Schewtschenko:2014fca} where the universal dependence (i. e., independent of the value of value of the cross section) for the IDM halo mass function was pointed out to exist (see also Ref.~\cite{Schewtschenko:2016fhl}, where the fit in Eq.~(\ref{eq:IDM}) appears to describe the halo mass function of all simulated models).  It is thus reasonable to assume that Eq.~(\ref{eq:IDM}) represents a good description of all the models considered in this work and it is the one we implement in our numerical calculations. Let us also emphasize that the fit to the halo mass function depends on the parameters describing the IDM transfer function $\mu$ and $\alpha_{\rm IDM}$ in $M_{\rm hm}$.

As both IDM and WDM scenarios result in a suppression of the small-scale matter power spectrum, it is possible to establish an approximate connection among these two schemes above the half-mode mass. By equating Eqs.~(\ref{eq:alphWDM}) and (\ref{eq:alphIDM}) one gets the correspondence
\begin{equation}
\left(\frac{\sigma_{\gamma \rm{DM}}}{\sigma_T}\right) \, \left(\frac{{\rm GeV}}{m_{\rm DM}}\right) \simeq 4.1 \times 10^{-9} \, \left( \frac{\text{keV}}{m_\text{WDM}} \right)^{2.4} ~,
\label{eq:W-Icorr-lin}
\end{equation}
which relates the DM-photon elastic scattering cross section in IDM scenarios to the DM mass in the WDM case that give rise to similar suppression of the linear power spectrum. This would help to map the constraints obtained in IDM scenarios to those corresponding to WDM scenarios, and viceversa, when scales not much smaller than the half-mode length are being tested. In our calculations, for IDM scenarios we focus on the range $\sigma_{\gamma {\rm DM}} \in [10^{-11} - 10^{-8}] \, \sigma_T \times \left(m_{\rm DM}/\textrm{GeV}\right)$, which according to the $m_{\rm DM}-\sigma_{\gamma {\rm DM}}$ relation above, corresponds roughly to $m_{\rm DM} \in [1-12]$~keV in WDM scenarios.

Notice, though, that Ref.~\cite{Schewtschenko:2014fca} shows the strong (qualitative and quantitative) similarities between the DM-neutrino and DM-photon interaction models. We will use those results to provide a rough estimate of the bounds on DM-neutrino interactions that could be obtained using the same method as the one presented here.

Also notice that Ref.~\cite{Schewtschenko:2014fca} showed indications that models with DM-neutrino interactions could be described in a very similar manner. In particular, from the performed simulations at $z=0$, the cases of $\sigma_{\gamma {\rm DM}} = 2.9 \times 10^{-9}\, \sigma_T \, (m_{\rm DM}/{\rm GeV})$ and of $\sigma_{\nu {\rm DM}} = 2.0 \times 10^{-9}\,\sigma_T \, (m_{\rm DM}/{\rm GeV})$ (for fixed half-mode mass) appear to result in very similar halo mass functions. We will consider this correspondence after deriving the bounds on IDM scenarios with DM-$\gamma$ elastic scattering.

\section{Ionization history of the universe}
\label{sec:reio}

As mentioned in the introductory section, the properties of the DM component could have a significant impact on the ionization history of the Universe. These effects can be exploited to constrain both the IDM and the WDM scenarios described above by studying the evolution of different reionization observables, such as the total ionized fraction $\bar{x}_i$ at different redshifts or the optical depth to reionization $\tau$ (see, e.g., Refs.~\cite{Barkana:2001gr, Sitwell:2013fpa, Bose:2016irl, Bose:2016hlz, Bose:2015mga, Yue:2012na, Somerville:2003sh, Yoshida:2003rm, Dayal:2014nva, Dayal:2015vca, Schultz:2014eia, Rudakovskiy:2016ngi, Lovell:2017eec} for previous analyses in this direction).

\subsection{Simulation and astrophysical parameters of the ionization history}

For the purpose of studying constraints from the ionization history of the Universe, we make use of the publicly available code {\tt 21cmFast}, based on excursion set formalism, perturbation theory and analytic prescriptions \cite{Mesinger:2010ne}. The code generates semi-analytic simulations of the evolved density, peculiar velocity, halo and ionization fields. While the main purpose of the code is the study of variations in the 21~cm signal due to changes in a given set of astrophysical and cosmological parameters, we use it here with the purpose of evaluating the ionization fraction evolution before and around the epoch of reionization (EoR) in different DM scenarios. We have adapted the default WDM implementation available in {\tt 21cmFast}, in particular in the definition of the transfer function and of the halo mass function, so as to account for the IDM and WDM descriptions provided in Section~\ref{sec:non-standard-dark}.

Let us  now describe the simplifying assumptions and ionization parameters considered in {\tt 21cmFast} to evaluate the ionized fraction. The total ionized fraction $\bar{x}_i$, is given by the covering factor of the fully ionized HII regions $Q_{\rm HII}$, plus a contribution from the averaged ionized fraction of the \emph{neutral} intergalactic medium (IGM), $x_e$~\cite{Mesinger:2012ys}
\begin{equation}
\label{eq:xetot}
 \bar{x}_i \simeq Q_{\rm HII}+(1-Q_{\rm HII}) \, x_e ~.
\end{equation}
Notice that the most relevant contribution to $\bar{x}_i$ at the epoch of reionization comes from $Q_{\rm HII}$. In {\tt 21cmFast}, the latter is characterized by\footnote{In {\tt 21cmFast} the different fields, including ionization, are resolved on a grid. Except for parameters such as $\zeta_{\rm UV}$ and $T_{\rm vir}$, described below, we use the {\tt 21cmFAST} default settings for our simulations corresponding to (200~Mpc)$^3$ comoving box with a 900$^3$ grid.}
\begin{equation}
Q_{\rm HII} = \frac{ \zeta_{\rm UV}f_{\rm coll}(>M_{\rm vir}^{\rm min})}{1-x_e} ~, 
\end{equation} 
where $\zeta_{\rm UV}$ is the UV ionization efficiency, which will be discussed below, and $f_{\rm coll}(>M_{\rm vir}^{\rm min})$ denotes the fraction of mass collapsed into halos with mass large enough to host star-forming galaxies (i.e., $M>M_{\rm vir}^{\rm min}$, see below). The latter is defined in terms of the halo mass function for a given cosmology $X$, introduced in Section~\ref{sec:non-standard-dark}, as
\begin{equation}
\label{eq:fcol}
 f_{\rm coll}^X (>M_{\rm vir}^{\rm min}) = \int_{M_{\rm vir}^{\rm min}} \, \frac{M}{\rho_{m,0}} \, \frac{dn^{X}}{dM} \, dM ~.
\end{equation}
On the other hand, the evolution of the local ionized fraction of the \emph{neutral} IGM, $x_e({\bf x}, z)$ results from
\begin{equation}
\label{eq:xe}
\frac{dx_e ({\bf x}, z)}{dz}  =  \frac{dt}{dz} \left(\Lambda_{\rm ion}
 - \alpha_{\rm A} \, C \, x_e^2 \, n_b \, \mathfrak{f}_{\rm H} \right) ~,  
\end{equation}
where $n_b = \bar{n}_{b, 0} (1+z)^3 (1 + \bar{\delta_b}({\bf x}, z))$ is the baryon number density, $\Lambda_{\rm ion}$ the ionization rate, $\alpha_{\rm A}$ the case-A recombination coefficient, $C\equiv \langle n_e^2 \rangle / \langle n_e \rangle^2$ is the clumping factor (set to one as default), with $n_e$ the electron number density, and $\mathfrak{f}_{\rm H} = n_{\rm H}/n_b$ is the hydrogen number fraction. Eq.~(\ref{eq:xe}) is solved numerically by means of the {\tt 21cmFAST} code, briefly described above. Once the ionization history is at hand, one can compute the optical depth to reionization $\tau$, defined as
\begin{equation}
\label{eq:tau} 
\tau=\sigma_T\int \bar{x}_i \, n_b \, dl ~,
\end{equation}
where $\sigma_T$ is the Thomson cross section and $dl$ is the line-of-sight proper distance.

Besides the DM parameters $m_{\rm DM}$ and $\sigma_{\gamma {\rm DM}}$, astrophysical parameters such as the ionization efficiency of UV photons, $\zeta_{\rm UV}$, and the minimum virial temperature, $T_{\rm vir}^{\rm min}$ (or equivalently the minimum virial mass $M_{\rm vir}^{\rm min}$, see below), have a strong impact on the evolution of $\bar{x}_i$.\footnote{The value of $\zeta_{\rm X}$, the number of X-ray photons per solar mass in stars (that represents the X-ray efficiency) has been fixed to $10^{56} \, M_\odot^{-1}$, which approximately corresponds to $N_{\rm X} \simeq 0.1$ X-ray photons per stellar baryon. Note that, if varied within an order of magnitude, in consistency with the (0.5-8)~keV integrated luminosity at $z = 0$~\cite{Mineo:2011id}, our results do not depend on the value of $\zeta_{\rm X}$. Nevertheless, a larger contribution from X-rays could have important consequences in more extreme scenarios~\cite{Mesinger:2012ys}.} It has already been pointed out that these parameters show degeneracies with the DM properties, affecting the small-scale matter power suppression (see, e.g., Refs.~\cite{Sitwell:2013fpa, Lopez-Honorez:2017csg}).  The UV ionizing efficiency $\zeta_{\rm UV}$ is assumed to be directly proportional to the fraction of ionizing photons escaping their host galaxy $f_{\rm esc}$, the number of ionizing photons per stellar baryons inside stars $N_\gamma$ and the fraction of baryons that form stars $f_\star$ (see, e.g., Ref.~\cite{Mesinger:2012ys}). Let us emphasize that $\zeta_{\rm UV}$ is assumed to be constant with redshift. The criterion encoded in the {\tt 21cmFast} code for a region to be considered ionized is
\begin{equation}
\zeta_{\rm UV} f_{\rm coll}(>M_{\rm vir}^{\rm min}) >1-x_e  ~.
\label{eq:ioncond}
\end{equation}
Here, we allow $\zeta_{\rm UV}$ to vary in the range $\zeta_{\rm UV} \in [5, 80]$ (see, e.g., Refs.~\cite{Liu:2015txa, Lopez-Honorez:2017csg} for previous analyses using similar ranges). Also, in Eqs.~(\ref{eq:fcol}) and~(\ref{eq:ioncond}), the minimum virial halo mass, $M_{\rm vir}^{\rm min}$, can be related to the threshold temperature for halos to host star-forming galaxies, $T_{\rm vir}^{\rm min}$, as~\cite{Barkana:2000fd}
 \begin{equation}
\label{eq:mminT}
M_{\rm vir}^{\rm min} (z) \simeq 10^8 \left(\frac{T_{\rm vir}^{\rm min}}{2 \times 10^4 \, {\rm K}} \right)^{3/2} \left(\frac{1+z}{10}\right)^{-3/2} M_\odot ~.
\end{equation}
The minimum value considered in this work is $T_{\rm vir}^{\rm min} = 10^4$~K, as lower temperatures have been shown to be insufficient to efficiently cool the halo gas through atomic cooling~\cite{Evrard:1990fu, Blanchard:1992, Tegmark:1996yt, Haiman:1999mn, Ciardi:1999mx}, while we take an upper limit of $T_{\rm vir}^{\rm min} \sim 2 \times 10^5$~K~\cite{Mesinger:2012ys, Greig:2015qca}. We take the same threshold temperature $T_{\rm vir}^{\rm min}$ for halos hosting ionizing and X-ray sources.

\begin{table}
	\begin{center}
	\resizebox{\textwidth}{!}{
		{\def\arraystretch{1.3}
			\begin{tabular}{|c||c|c|c|c|c|} 
				\hline
				& \, $\alpha_X$  [Mpc/$h$] \, & \, $M_{\rm hm} \, [M_\odot]$ \, & \, $\zeta_{\rm UV}$ \, & \, $T_{\rm vir}^{\rm min}$ [K]  \, & \, $\tau$ \, \\
				\hline				
				$\sigma_{\gamma \rm{DM}} = 6.3\times 10^{-10} (\sigma_T \times {m_{\rm DM}}/{\rm GeV})$ &  \multirow{2}{*}{ 0.0071} & \multirow{2}{*}{$6.9  \times 10^8$} & \multirow{2}{*}{55} & \multirow{2}{*}{$10^5$} & 0.061 \\
				$m_{\rm{WDM}} = 2.15$ keV & & & & & 0.059  \\
				\hline
				$\sigma_{\gamma \rm{DM}}  = 7.9\times 10^{-11} (\sigma_T \times {m_{\rm DM}}/{\rm GeV})$ &  \multirow{2}{*}{ 0.0020} & \multirow{2}{*}{$3.5  \times 10^7$} & \multirow{2}{*}{30} & \multirow{2}{*}{$5 \times 10^4$} & 0.064 \\
				$m_{\rm{WDM}}$ = 5.17 keV & & & & & 0.063  \\
				\hline
			\end{tabular}
		}
		}
		\caption{IDM and WDM benchmark models, corresponding to the same $M_{\rm hm}$. See Figs.~\ref{fig:ioWDMIDM} and~\ref{fig:Tb} for the ionization histories and the 21~cm signals expected for these two models.}
		\label{tab:tabbench}
		
	\end{center}
\end{table}

\begin{figure}[t]
\centering
\includegraphics[width=0.8\textwidth]{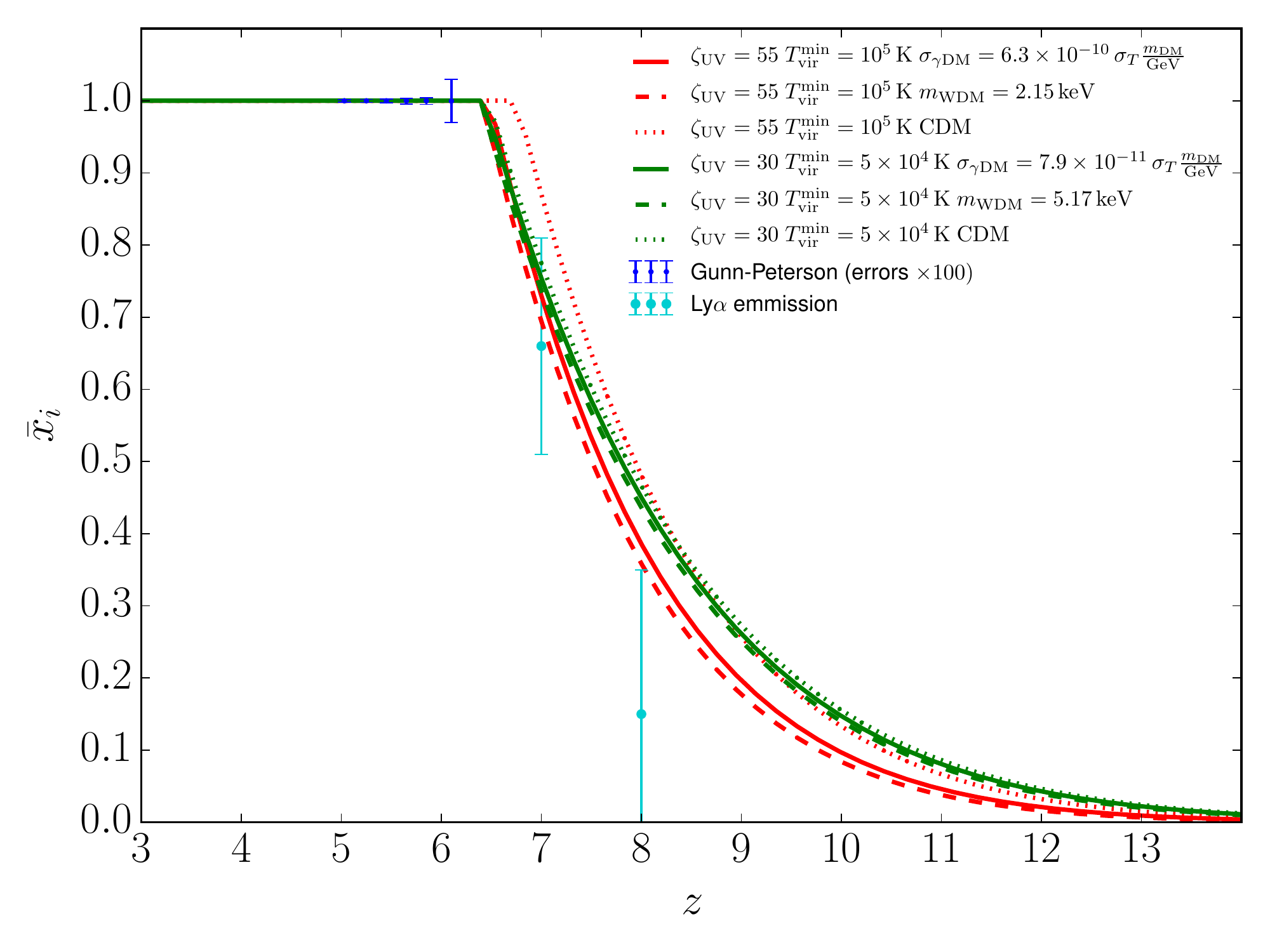} 
\caption{Ionization history for IDM (solid curves), WDM (dashed curves) and CDM (dotted curves) scenarios for two combinations of values of the ionization efficiency and the minimum virial temperature: $\zeta_{\rm UV} = 30$ and $T_{\rm vir}^{\rm min} = 5 \times 10^4$~K (green curves), and $\zeta_{\rm UV} = 55$ and $T_{\rm vir}^{\rm min}= 10^5$~K (red curves). The parameters of the IDM and WDM scenarios are also indicated (see Tab.~\ref{tab:tabbench}). The blue and cyan points and the associated error bars correspond to the low- and high-redshift $\bar x_i$ measurements considered in Section~\ref{sec:reio} (see Tab.~\ref{tab:tab1}). Notice that the errors associated to Gunn-Peterson effect measurements (in blue) have been magnified by a factor of 100 to be visible.}
\label{fig:ioWDMIDM}
\end{figure}

For illustrative purposes, in Fig.~\ref{fig:ioWDMIDM} we show the resulting ionization history for different benchmark scenarios (see Tab.~\ref{tab:tabbench}), which are allowed at $95\%$ confidence level (CL), to be discussed in Section~\ref{sec:comb-constr}: IDM (solid curves), WDM (dashed curves) and CDM (dotted curves), for two values of the ionization efficiency and of the minimum virial temperature: $\zeta_{\rm UV} = 30$ with $T_{\rm vir}^{\rm min}= 5 \times10^4$~K (green curves) and $\zeta_{\rm UV} = 55$ with $T_{\rm vir}^{\rm min} = 10^5$~K (red curves). The chosen values for the DM-photon elastic scattering cross section and the WDM mass are such that the half-mode mass is the same in the IDM and WDM scenarios. As can be seen from the figure (solid and dashed curves of each color), this results into very similar ionization histories, although WDM scenarios give rise to a slightly more delayed reionization than in the corresponding IDM scenarios. This is related to the extra amount of power at small scales in the IDM case with respect to the WDM case due to the oscillating matter power spectrum at scales smaller than $\lambda_{\rm hm}$. As mentioned above, this effect is at the origin of the slightly modified description of the halo mass function of the IDM scenario in Eq.~(\ref{eq:IDM}), compared to the WDM case in Eq.~(\ref{eq:WDM})~\cite{Schewtschenko:2014fca}. All in all, distinguishing these two scenarios using this type of observables would be a very challenging task (if possible at all).

\subsection{Ionization history measurements}

To jointly constrain the ionization history of the Universe and the DM properties, we exploit two sets of observables. First of all, we use measurements of the optical depth from the last-scattering surface to reionization, $\tau$, which provides information on the integrated ionization history of the Universe and affects the CMB photon temperature and polarization spectra. Indeed, the most accurate measurements of $\tau$ are obtained by means of CMB data and consequently, for our numerical analyses, we impose a Gaussian prior on $\tau$ using the Planck measurement $\tau = 0.055 \pm 0.009$~\cite{Aghanim:2016yuo}. We first calculate the redshift evolution of the total ionized fraction, $\bar{x}_i (z)$, using the {\tt 21cmFAST} code~\cite{Mesinger:2007pd, Mesinger:2010ne}, as previously explained. Then, we determine the value of $\tau$ for each of the models studied here by means of the Boltzmann solver code CAMB (Code for Anisotropies in the Microwave Background)~\cite{Lewis:1999bs}, that has been modified to allow for any possible ionization history $\bar{x}_i (z)$, including those corresponding to WDM or IDM scenarios.

Different astrophysical observations can provide very valuable information on the evolution of the global ionization fraction. At low redshifts, measurements of the Gunn-Peterson optical depth from bright quasars at six different redshifts, $z = 5.03, \, 5.25, \, 5.45, \, 4.65, \, 5.85$, $6.10$~\cite{Fan:2005es}, as well as the distribution of dark gaps in quasar spectra at $z=5.6$ and $z = 5.9$~\cite{McGreer:2014qwa}, indicate that reionization has to be completed by $z \sim 6$. We consider this {\it low-$z$} set of measurements to further constrain the different model parameters.\footnote{Notice that in our previous analyses~\cite{Lopez-Honorez:2017csg}, we did not consider the $95\%$~CL constraints from the low redshift data set. Instead, we considered these measurements as lower limits.}

On the other hand, we also use data from Ly$\alpha$ emission in star-forming galaxies at earlier times ($z \gtrsim 7$). Extrapolating the behavior at lower redshifts~\cite{Santos:2003pc, Malhotra:2004ef, McQuinn:2007dy, Mesinger:2007jr, Stark:2010qj, Stark:2010nt, Fontana:2010ms, Dijkstra:2011de, Pentericci:2011ft, Ono:2011ew, Caruana:2012ww, Treu:2013ida, Caruana:2013qua, Tilvi:2014oia}, these results indicate that reionization is not complete at those epochs. Concretely, in this work, we consider data at $z = 7$ and $z = 8$~\cite{Schenker:2014tda}, which use the models of Ref.~\cite{McQuinn:2007dy}, and we refer to them as the {\it high-$z$} data set. A compilation of all data can be found in Ref.~\cite{Bouwens:2015vha}.

Therefore, in practice, we compute three different $\chi^2$ functions, one for each type of data ($\tau$ measurements, Gunn-Peterson optical depth and dark gaps in quasar spectra data, and high-$z$ constraints from Ly$\alpha$ emission) and then we add them up. Tab.~\ref{tab:tab1} shows the results of these measurements, together with their associated errors, for the Gunn-Peterson and high-redshift Ly$\alpha$ emission.

\begin{table}
\begin{center}
{\def\arraystretch{1.4}
\resizebox{\textwidth}{!}{
\begin{tabular}{|c||c|c|c| } 
 \hline
 \, Data \, & \, Redshift \, & \, $x_{\rm HII}$ \, & \, Reference \, \\
 \hline
 \hline
 \multirow{6}{*}{Gunn-Peterson effect}
 & 5.03 & \, 0.9999451$_{-0.0000165}^{+0.0000142}$ \, & \multirow{6}{*}{\cite{Fan:2005es}} \\[1ex] 
 & 5.25 & 0.9999330$_{-0.0000244}^{+0.0000207}$ &  \\[1ex]
 & 5.45 & 0.9999333$_{-0.0000301}^{+0.0000247}$  &  \\[1ex]
 & 5.65 & 0.9999140$_{-0.0000460}^{+0.0000365}$  &  \\[1ex]
 & 5.85 & 0.9998800$_{-0.0000490}^{+0.0000408}$  &  \\[1ex]
 & 6.10 & 0.99957$\pm$0.00030 &  \\
 \hline
 \multirow{2}{*}{Dark gaps in quasar spectra}
 & 5.6 & $>$0.91 & \multirow{2}{*}{\cite{McGreer:2014qwa}} \\[1ex] 
 & 5.9 &  $>$0.89 &  \\
 \hline
 \multirow{2}{*}{\, Ly$\alpha$ Emission in Galaxies (High redshift) \, }
 & 7 & 0.66 $\pm$ 0.15 & \multirow{2}{*}{\cite{Schenker:2014tda}} \\[1ex] 
 & 8 & 0.15 $\pm$ 0.20 & \\ 
 \hline
\end{tabular}
}
}
\label{tab:tab1}
\caption{Set of $\bar x_i$ data used in this work (see also Ref.~\cite{Bouwens:2015vha}).}
\end{center}
\end{table}

\section{Number of Milky Way satellite galaxies}
\label{sec:Nsat}

As previously discussed, the standard CDM framework may be facing a problem at subgalactic scales, the so-called ``missing satellite problem''. More specifically, the current number of observed satellite galaxies is $N_{\rm gal} \sim 50$~\cite{Drlica-Wagner:2015ufc} and extrapolations to the entire MW virial volume show that $N_{\rm gal} \sim 150$~\cite{Newton:2017xqg} are expected to be present, while CDM N-body simulations~\cite{Springel:2008cc, Griffen:2016ayh, Garrison-Kimmel:2013eoa} predict that there should be $N_{\rm sub} \sim 1000$ dark subhalos with $M_{\rm sub} > 10^{7} \, M_\odot$. Nevertheless, this can be explained by the suppression of galaxy formation efficiency for low mass subhalos~\cite{Behroozi:2012iw, Brooks:2012vi, Moster:2012fv}. Therefore, the number of observed satellite galaxies does not seem to be in contradiction with numerical DM-only simulations, once the effects of baryons are included. Indeed, we will use these arguments to set a bound on the maximum amount of suppression which is allowed by current satellite counts. In what follows, we shall review the observational status and our treatment of the MW satellite galaxy counts.

\subsection{Observational Status}

The current number of discovered satellite galaxies in the MW is 54, out of which 11 are the so-called classical ones, 17 have been discovered by DES~\cite{Bechtol:2015cbp, Drlica-Wagner:2015ufc}, 17 by SDSS~\cite{Ahn:2012fh, Koposov:2009ru} and 9 have been found in other surveys (see Appendix A of Ref.~\cite{Newton:2017xqg} for a comprehensive catalog). Extrapolation to full sky of the total number of dwarf satellite galaxies in the MW has been a subject of intense study~\cite{Koposov:2007ni, Tollerud:2008ze, Hargis:2014kaa, Newton:2017xqg}. We shall make use of the latest estimation~\cite{Newton:2017xqg}, which accounts for the latest DES discoveries and newer simulations, and leads to the constrain $N_\text{gal} > 85$ at $95$\% ~CL across the entire sky. Finally, it is worth to note that the estimate of the number of satellites in the MW from Ref.~\cite{Newton:2017xqg} has been obtained from radial extrapolations of the subhalo distribution from CDM simulations. Nevertheless, the radial subhalo distribution has been shown to be fairly universal and independent of the DM properties~\cite{Bose:2016irl}. Therefore, in the following, we will apply the bound
\begin{equation}
 N_\text{gal} > 85 \quad \textrm{at 95\% CL} ~.
\end{equation}

\subsection{Number of subhalos in IDM scenarios}
\label{subsec:N_sat}

In order to set constraints on the IDM cross section, we now compute the number of satellite galaxies in the MW in terms of $\sigma_{\gamma{\rm DM}}$. The most sophisticated and accurate way of performing this calculation is through N-body simulations, and this approach has been followed in the past for IDM~\cite{Boehm:2014vja, Schewtschenko:2015rno} (see also Refs.~\cite{Lovell:2013ola, Jethwa:2016gra} for the WDM case). Here, instead, we follow an analytical approach along the same lines of the recent Ref.~\cite{Kim:2017iwr}, circumventing the computationally expensive N-body simulations. This method requires as inputs the subhalo mass function for IDM cosmologies, and the probability for a subhalo to host a galaxy, $f_\text{lum}(M)$. The number of subhalos is defined as~\cite{Kim:2017iwr}
\begin{equation}
N_\text{sub} = \int_{M_\text{min}}^{{M}_\text{host}} \frac{d N}{dM} \, d M ~,
\label{eq:N_sub}
\end{equation}
where $M_\text{min}$ is the minimum subhalo mass considered, $M_\text{host}$ is the mass of the host galaxy (in our case the MW) and $dN/dM$ is the subhalo mass function. To obtain the number of satellite galaxies (namely, luminous subhalos) a correction to the above expression is required, so that the probability that a subhalo of a given mass hosts a luminous galaxy, $f_\text{lum}(M)$, is accounted for and hence~\cite{Kim:2017iwr},
\begin{equation}
N_\text{gal} = \int_{{M}_\text{min}}^{{M}_\text{host}} \frac{d N}{dM} \, f_\text{lum} (M) \, d M \,.
\label{eq:N_lum}
\end{equation}

From N-body simulations within the CDM scenario, a fit to the subhalo mass function has been obtained in Ref.~\cite{Dooley:2016xkj},
\begin{equation}
\frac{d N^{\rm  CDM}}{dM^{\rm peak}} = K_0 \, \left(\frac{M^{\rm peak}}{M_\odot}\right)^{-\chi} \, \frac{M_\text{host}}{M_\odot} ~,
\label{eq:CDM_sub}
\end{equation}
where $K_0 = 1.88 \times 10^{-3} \, M_\odot^{-1}$ and $\chi = 1.87$ are fitting parameters and $M^{\rm peak}$ represents $M_{200}$ at peak (i.e., the maximum mass of a subhalo achieved over its history enclosed by a volume that is $200 \, \rho_c$, with $\rho_c$ the critical density of the Universe). Note that the number of subhalos also depends on the mass of the host halo. In the particular case of the MW, the range of values for the MW mass~\cite{Wang:2015ala} introduces an additional uncertainty, which we take into account. 

For the case of IDM scenarios, to the best of our knowledge, the only available N-body simulations are those from Ref.~\cite{Boehm:2014vja}, but unfortunately they did not provide a fit to the IDM subhalo mass function. Nevertheless, WDM simulations have shown that the suppression in the halo mass~\cite{Schneider:2011yu} and subhalo~\cite{Lovell:2013ola} mass functions are fairly similar, agreeing within 40\%. Furthermore, the subhalo suppression is always more pronounced, as expected for non-CDM scenarios with reduced power at small scales. Therefore, following a conservative approach, for the IDM subhalo mass function we use the same parametric form of suppression, with respect to CDM, as given for halos in Eq.~(\ref{eq:IDM}), namely,
\begin{equation}
\frac{d N}{dM}^{\rm  IDM} = \left(1 + \frac{M_{\rm hm}}{b \, M}\right)^{a} \left(1 + \frac{M_{\rm hm}}{g \, M}\right)^{c}  \frac{d N}{dM}^{\rm  CDM} ~,
\label{eq:IDM_sub}
\end{equation}
with $a = -1$, $b = 0.33$, $g = 1$, $c = 0.6$ and $M$ is the mass at $z = 0$, which is different from $M^{\rm peak}$ due to the effect of tidal stripping. They are related with the scaling relation $(M/M_\odot) = (M^{\rm peak}/M_\odot)^{0.965}$, which we obtain from a fit to the high-resolution halo catalog of the ELVIS simulation~\cite{Garrison-Kimmel:2013eoa}.

This is a conservative approach, as adopting this suppression should lead to a larger number of subhalos/galaxies than those resulting from dedicated IDM simulations. For the sake of comparison between IDM and WDM scenarios, we use of the following description of the WDM subhalo mass function~\cite{Lovell:2013ola}: 
\begin{equation}
\frac{d N}{dM}^{\rm  WDM} = \left(1 + g_s \frac{M_{\rm hm}}{M} \right)^{-b_s} \frac{d N}{dM}^{\rm  CDM} ~,
\label{eq:WDM_sub}
\end{equation}
where $g_s = 2.7$, $b_s = 0.99$. 

Assuming all DM subhalos to host luminous galaxies would largely overestimate the number of visible satellites~\cite{Bullock:2000wn, Somerville:2001km, Benson:2001at}. This is the reason why we have introduced the function $f_\text{lum}$ in Eq.~(\ref{eq:N_lum}). We shall follow the results of Ref.~\cite{Dooley:2016xkj} for its description, bearing in mind they were obtained from a CDM merger history. In particular, the fraction of DM halos that host luminous galaxies, $f_\text{lum}$, depends on the ionization model. Indeed, UV photons able to ionize hydrogen prevent sufficient cooling and gas accretion for star formation. The importance of this effect is actually encapsulated in the parameter $T_{\rm vir}^{\rm min}$, introduced in Section~\ref{sec:reio}. Here, we directly use the results of Ref.~\cite{Dooley:2016xkj} that explored possible scenarios with reionization redshifts $z_\text{re} = 14.4$, $11.3$ and $9.3$. Their most conservative choice, giving rise to the largest number of satellite galaxies, is a model with a reionization redshift of $z_\text{re} = 9.3$, and we focus on this particular realization.\footnote{A higher reionization redshift reduces the probability for a halo of reaching the critical size for $H_2$ and atomic cooling before reionization, reducing the probability for star-formation in such a halo and therefore suppressing the final number of satellite galaxies. In this regard, our numbers are conservative as the reionization redshift in this work is lower than the one used in Ref.~\cite{Dooley:2016xkj} to obtain $f_{\rm lum}$.} The form of $f_\text{lum}$ is a fast-rising function at a particular value of the subhalo mass (close to a step-function) which depends on the reionization redshift, reaching $f_\text{lum} \sim 0.5$ at $M \sim 2 \times 10^8 \, M_\odot$ for $z_\text{re} = 9.3$ (see Fig.~12 of Ref.~\cite{Dooley:2016xkj}).

In the next section, we use the results above in order to constraint IDM scenarios. In Section~\ref{sec:gal-cons}, we also provide a direct comparison between the results of the analytic approach followed here and those obtained in Ref.~\cite{Boehm:2014vja}, based on IDM simulations. They are consistent with each other, which establishes the validity of the method we follow.

\section{Results and Prospects}
\label{sec:results}

\subsection{Reionization constraints}
\label{sec:reio-cons}

\begin{figure}[t]
	\includegraphics[width=0.32\textwidth]{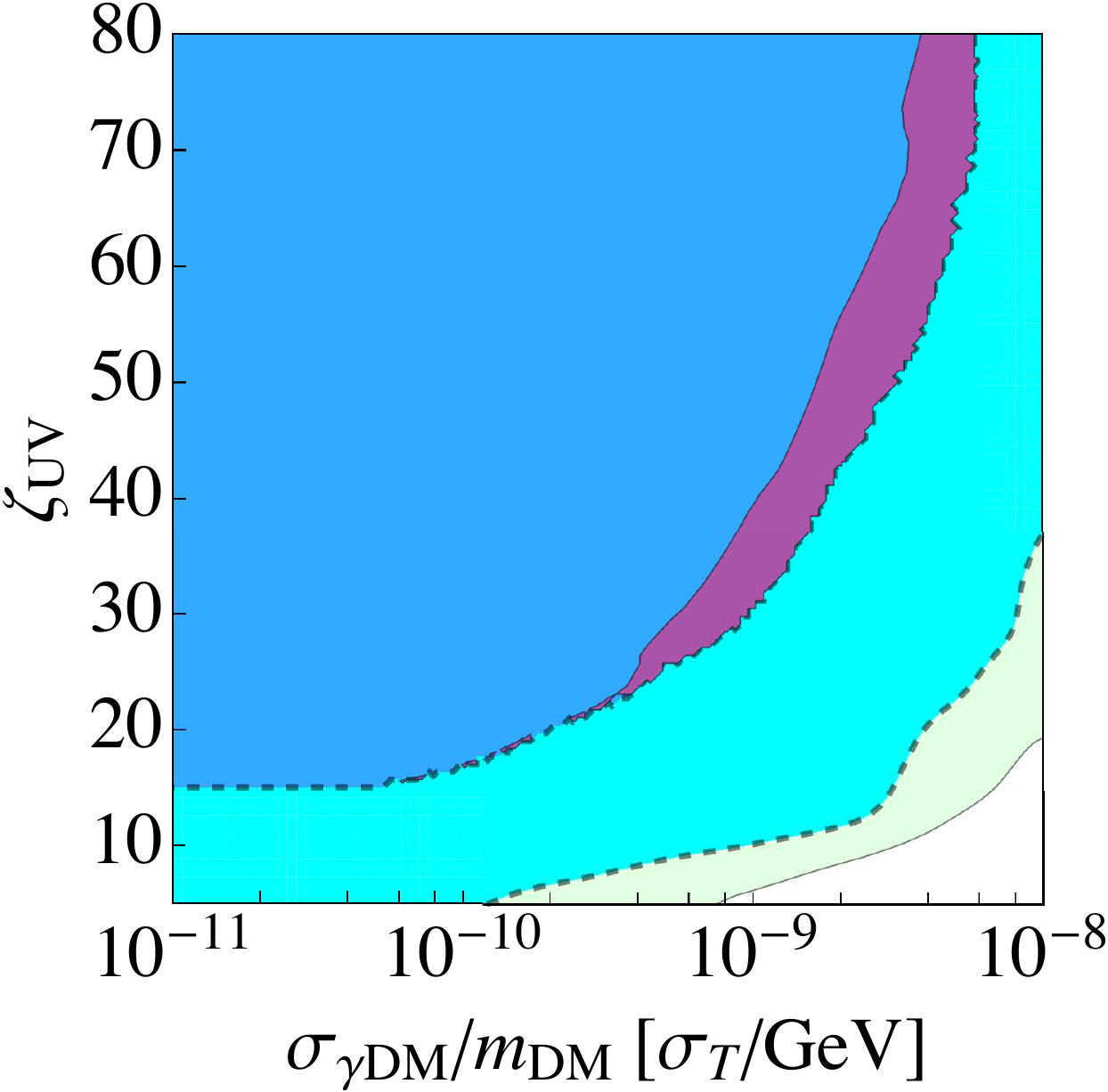}
	\includegraphics[width=0.32\textwidth]{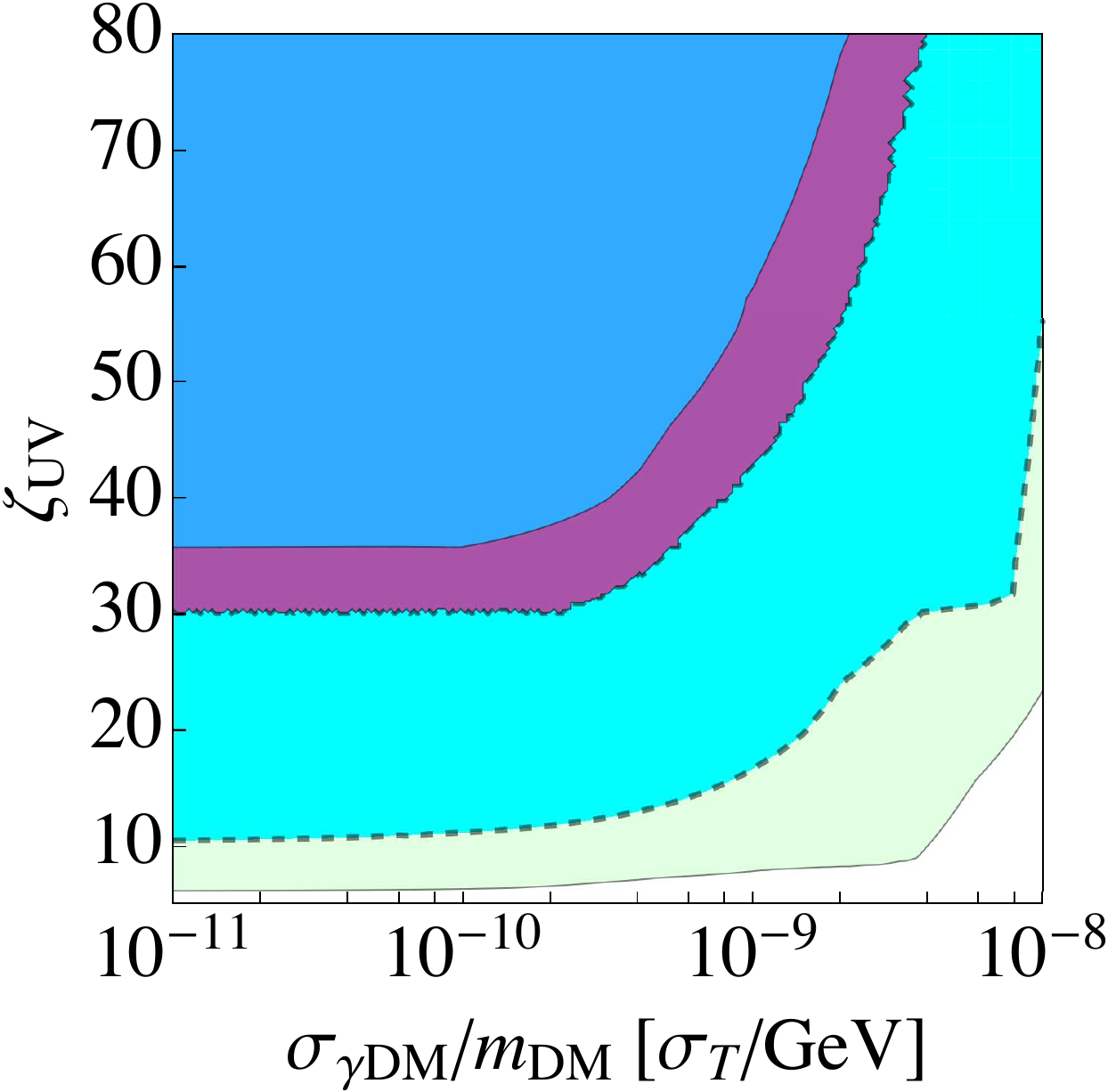} 
	\includegraphics[width=0.32\textwidth]{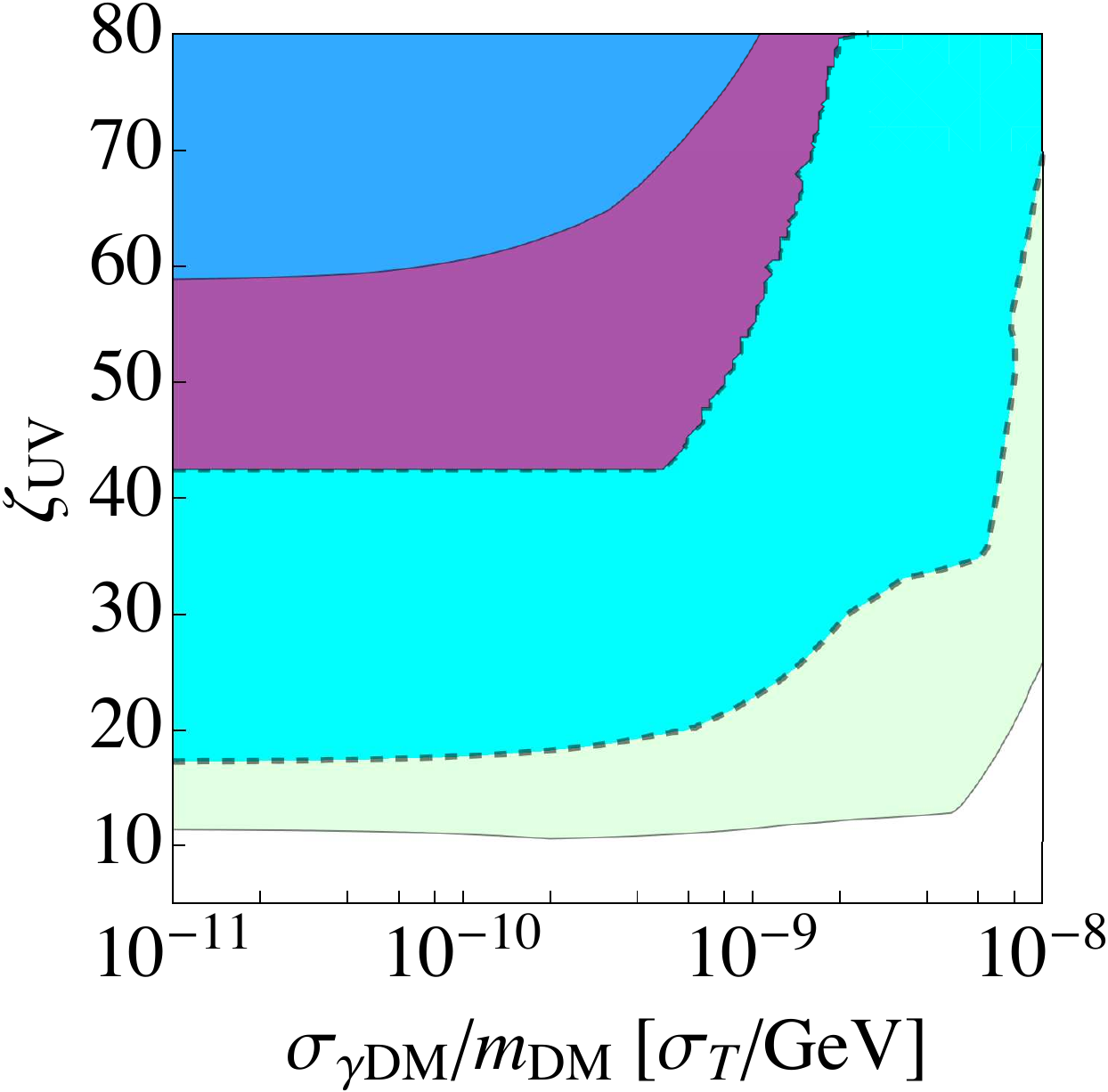} 
	\caption{Constraints at $95\%$~CL in the $\left(\sigma_{\gamma {\rm DM}}/m_{\rm DM}, \, \zeta_{\rm UV}\right)$ plane, from $\tau$ and $\bar x_i$ measurements, for $T_{\rm vir}^{\rm min} =10^4$, $5 \times 10^4$ and $10^5$~K (from left to right).  The cyan (dark blue) contours denote the $95\%$~CL allowed regions from high-$z$ (low-$z$) $\bar x_i$ data. The light green regions below these contours represent the $95\%$~CL constraints from $\tau$ measurements. The dark purple regions denote the joint constraints from all three measurements.}
	\label{fig:Chi2xe}
\end{figure}

The results from the analyses of the ionization history data are depicted in Fig.~\ref{fig:Chi2xe}, where we show the $95\%$~CL bounds arising from the three different $\chi^2$ analyses in the $\left( (\sigma_{\gamma {\rm DM}}/m_{\rm DM}), \,  \zeta_{\rm UV}\right)$ plane. We show one contour for each type of data ($\tau$ measurements in light green, Gunn-Peterson optical depth plus dark gaps in quasar spectra data at low-$z$ in blue, and high-$z$ constraints from Ly$\alpha$ emission in cyan). We also show the combined $95\%$~CL region in dark purple. From left to right, the three panels illustrate the constraints for a value of the minimum virial temperature of $T_{\rm vir}^{\rm min} =$ $10^4$, $5 \times 10^4$ and $10^5$~K.

First of all, notice that the $\tau$ measurement, being an integrated quantity, can only set modest constraints on the parameter space, as can be seen from the green contours in Fig.~\ref{fig:Chi2xe} (see also Ref.~\cite{Lovell:2017eec} for a related discussion). Furthermore, the larger the minimum virial temperature, the worse its constraining power, as even if larger values of $T_{\rm vir}^{\rm min}$ imply lower values of $\tau$, this effect can be easily compensated with a higher ionization efficiency, $\zeta_{\rm UV}$. Even for the lowest $T_{\rm vir}^{\rm min}=10^4$~K we consider (left panel), a larger value of $\sigma_{\gamma \textrm{DM}}$ (i.e., a more important suppression of power at small scales that would delay reionization) could be compensated by a larger UV ionizing efficiency $\zeta_{\rm UV}$. This shows the strong (positive) correlation of these two parameters. Indeed, a similar correlation appears for the high-$z$, Ly$\alpha$ emission data (cyan contours) for all values of $T_{\rm vir}^{\rm min}$: the smaller $T_{\rm vir}^{\rm min}$ (i.e., the smaller the minimum mass for star formation and thus the earlier the period of reionization), the larger the value of $\zeta_{\rm UV}$ needed to recover consistency with data.

In addition, with the low-$z$ data set, which indicates $\bar x_i\simeq 1$ with a high degree of precision (see Tab.~\ref{tab:tab1}), one can further constrain the parameter space (see dark blue regions in Fig.~\ref{fig:Chi2xe}). In order to explain these data, a large ionization efficiency $\zeta_{\rm UV} > 30$ is required to ensure that reionization is complete by $z \simeq 5 - 6$. As it might not be possible to compensate the delayed reionization caused by a large DM-photon cross section due to the suppression of power at small scales, this results into an upper bound on the cross section, $\sigma_{\gamma \rm{DM}}/\sigma_T \lesssim  4\times 10^{-9} \left( {m_{\rm DM}}/{\rm GeV}\right)$ at $95\%$~CL, for all values of $T_{\rm    vir}^{\rm min}$. Indeed, it appears that low and high-$z$ data sets constrain different regions of the parameter space. Therefore, when these data sets are combined (see the dark purple contours in Fig.~\ref{fig:Chi2xe}),  upper (lower) bounds on the interacting DM cross section (UV efficiency) can be obtained. We would like to stress this point, as the combination of more precise low-$z$ and high-$z$ measurements of the ionization history could set strong limits to IDM scenarios or to  any other non-CDM cosmologies. In this regard, 21~cm probes, which are expected to map the HI along the Universe's history and therefore cover a wide redshift range, may help in testing this type of scenarios (see Sec.~\ref{sec:21cm}).

\begin{figure}[t]
	\centering
	\includegraphics[width=0.75\textwidth]{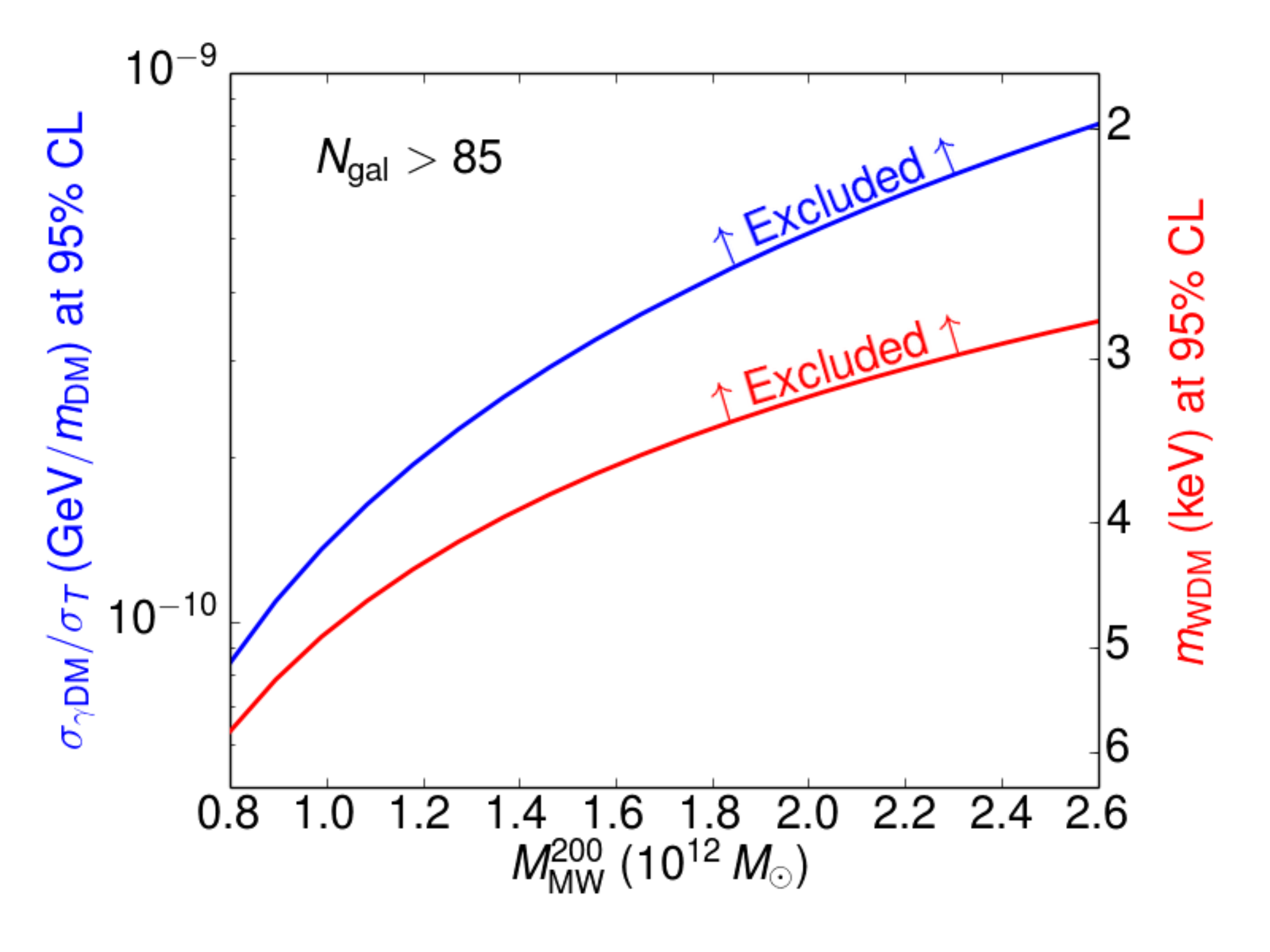} 
	\caption{The top blue (bottom red) curve shows the $95\%$~CL upper (lower) limits on the DM-photon cross section (WDM mass) as a function of the MW mass. The $y$-axes are calibrated so that the models have the same transfer function up to the first damped oscillatory mode in the IDM and WDM matter power spectra (see Eq.~(\ref{eq:pwdm}) and Fig.~\ref{fig:transfer}).}
	\label{fig:Sat_res}
\end{figure}

\subsection{Milky Way satellites constraints}
\label{sec:gal-cons}

We focus now on the constraints from MW satellite galaxies. By means of Eq.~(\ref{eq:N_lum}) with the CDM subhalo mass functions, and using the IDM subhalo suppression, Eq.~(\ref{eq:IDM_sub}), we have calculated the number of satellite galaxies in the MW as a function of the photon-DM elastic scattering cross section. Imposing the aforementioned $95\%$~CL lower limit in the number of satellite galaxies (i.e., $N_\text{gal} > 85$~\cite{Newton:2017xqg}), we obtain the bounds on the IDM cross section depicted by the blue curve in Fig.~\ref{fig:Sat_res}. Notice that we show the results as a function of the MW mass $M_{\rm MW}^{200}$, within its expected mass range~\cite{Wang:2015ala}. The most conservative $95\%$~CL upper limit on the IDM cross section is found for the highest value of $M_{\rm MW}^{200}$ considered here (i.e., $(\sigma_{\gamma \rm{DM}}/\sigma_T) < 8 \times 10^{-10} \, \left( {m_{\rm DM}}/{\rm GeV}\right)$). For the sake of completeness, we also show the $95\%$~CL lower bounds on the WDM mass by the red curve in Fig.~\ref{fig:Sat_res}. The most conservative $95\%$~CL lower limit corresponds to $m_{\rm WDM} > 2.8$~keV (for a thermal candidate). Notice that these results imply an order of magnitude improvement on the DM-photon elastic scattering cross section over those previously obtained in Ref.~\cite{Boehm:2014vja}, while the bound on the WDM mass we find is very similar to the results from N-body simulations from Ref.~\cite{Jethwa:2016gra}.

\begin{figure}[t]
	\centering
	\includegraphics[width=0.75\textwidth]{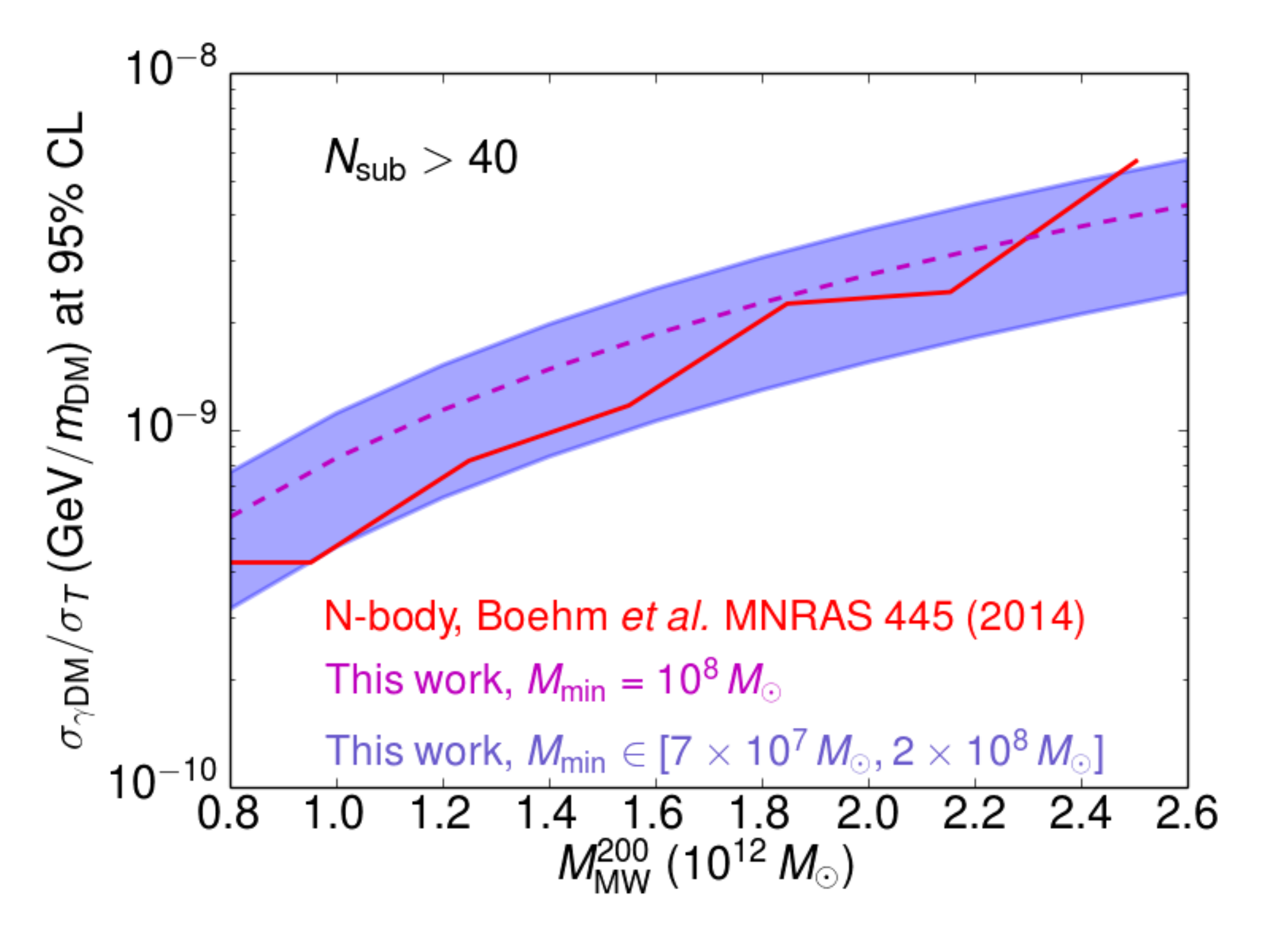} 
	\caption{Upper limits on the DM-photon elastic scattering cross section as a function of the MW mass, obtained at $95\%$~CL. They are calculated following the method described in Sec~\ref{subsec:N_sat} by setting $N_\text{sub} > 40$, considering subhalos with masses above the range $M_\text{sub}^\text{min} \in [7 \times 10^7, \, 2 \times 10^8]~M_\odot$ (blue contour) and also by fixing the subhalo mass to $M_\text{sub} \sim 10^8\, M_\odot$ (dashed curve). For comparison purposes, we also report  the constraints from N-body simulations from Ref.~\cite{Boehm:2014vja} with the red solid line.}
	\label{fig:comp_IDM}
\end{figure}

Let us now compare our results with those of previous analyses of MW satellite galaxies.  While a direct comparison of the bounds from MW number counts derived here to the ones obtained in Ref.~\cite{Boehm:2014vja} is non-trivial, we nevertheless make a comparison to clearly state the validity of our method. In Ref.~\cite{Boehm:2014vja}, the cumulative number count of MW satellites was studied as a function of their maximal circular velocity, $V_\text{max}$, the latter being a measure of their mass. In contrast, our method relies on the subhalo mass $M$. This implies that, for a direct comparison, we would need a relationship between $V_\text{max}$ and $M$ for IDM, which is still missing in the literature. Furthermore, due to the finite resolution, Ref.~\cite{Boehm:2014vja} only considered subhalos with a maximum peak velocity $V_\text{max} \gsim 8 \, \text{km/s}$, while there are known satellites with $V_\text{max} \sim 5.7 \, \text{km/s}$~\cite{Wolf:2009tu}.  We know, however, from WDM simulations~\cite{Lovell:2013ola} (that show a similar matter power suppression to that of IDM) that $V_\text{max} = 8 \, \text{km/s} $ corresponds to $M_\text{sub} \sim 10^8\, M_\odot$. As a result, the total number of satellites with $V_\text{max} > 8 \, \text{km/s}$ is expected to be $N_\text{sub} (V_\text{max} > 8 \, \text{km/s}) > 40$ at $95\%$~CL~\cite{Willman:2009dv}. Finally, since the simulations in Ref.~\cite{Boehm:2014vja} only considered DM, they did not include the luminosity function of subhalos discussed above. This is the reason for using here the fraction of DM halos that host luminous galaxies $f_{\rm lum}(M)$ from CDM simulations (see Section~\ref{subsec:N_sat} for the derivation of the constraints in Section~\ref{sec:gal-cons}).

All in all, in order to make a fair comparison with previous estimates, \textit{(a)} we shall compare subhalos and not galaxies; \textit{(b)} we shall require that the minimum mass of the subhalos to be considered satisfies $M_\text{sub}^\text{min} \sim 10^8\, M_\odot$; and \textit{(c)} we should impose $N_\text{sub} > 40$. The comparison of our results to those of Ref.~\cite{Boehm:2014vja} under these conditions is illustrated in Fig.~\ref{fig:comp_IDM}. The red dashed curve shows our results for a minimum subhalo mass $M_\text{sub} \sim 10^8\, M_\odot$, see Eq.~(\ref{eq:N_lum}), which roughly corresponds to $V_\text{max} = 8$~km/s. We also illustrate with the blue region the results we would obtain for a minimum subhalo mass within the interval $M_\text{sub}^\text{min} \in [7 \times 10^7, \, 2 \times 10^8]~M_\odot$. This blue region should be compared with the results based on IDM N-body simulations previously obtained in Ref.~\cite{Boehm:2014vja}, shown here with the red solid line. As the agreement between the two approaches is fairly good, the method used here to derive bounds on the IDM cross section using MW number counts is well justified.

\subsection{Combination of constraints}
\label{sec:comb-constr}

\begin{figure}[t]
	\centering
	\includegraphics[width=0.65\textwidth]{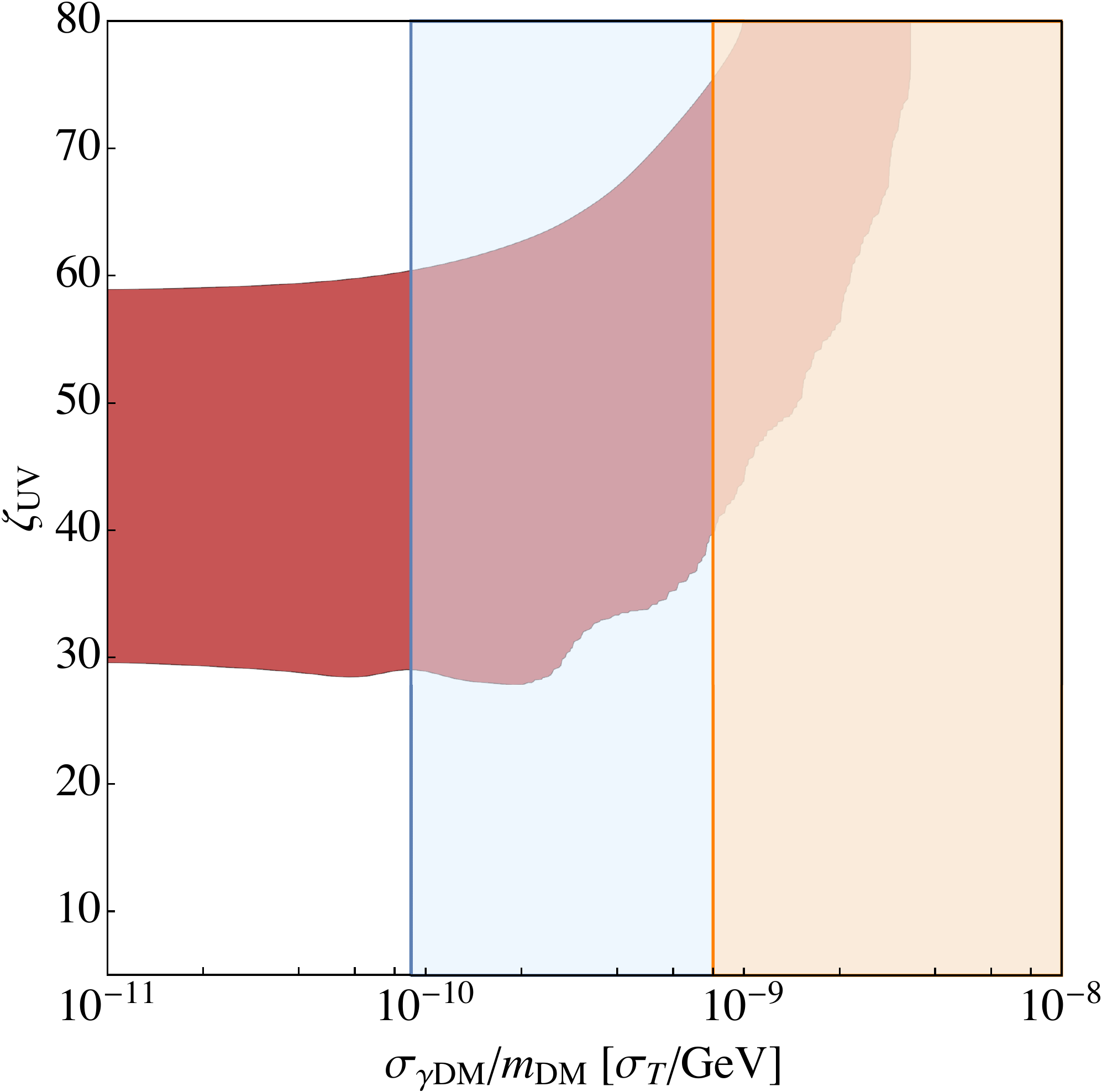}
	\caption{Global constraints, at $95\%$~CL, on the $\left(\sigma_{\gamma \textrm{DM}}/m_{\rm DM}, \, \zeta_{\rm UV}\right)$ plane. We depict the allowed region from the combination of $\tau$ and $\bar x_i$ data, after profiling over $T_{\rm vir}^{\rm min}$ (red contour); and the excluded region from satellite counts, assuming a MW mass of $ M_{\rm MW}^{200} = 0.8 \times 10^{12}~ M_{\odot}$ (shaded blue region) and $ M_{\rm MW}^{200} = 2.6 \times 10^{12}~ M_{\odot}$ (shaded orange region).} 
	\label{fig:Chi2xet}
\end{figure}

In Fig.~\ref{fig:Chi2xet} we depict the $95\%$~CL bounds from the global $\chi^2$ analysis (profiling over $T_{\rm vir}^{\rm min}$), by combining low-$z$ (Gunn-Peterson and dark gaps in quasar spectra data) and high-$z$ $\bar x_i$ measurements together with the constraints on the reionization optical depth $\tau$ from the Planck satellite, in the $\left((\sigma_{\gamma \textrm{DM}}), \, \zeta_{\rm UV}\right)$ plane. Notice the large degeneracy between the DM-photon cross section and the UV efficiency, already discussed in Section~\ref{sec:reio-cons} and also present in Fig.~\ref{fig:Chi2xe}. We have superimposed the allowed regions from the MW satellite number counts likelihood. Note that for our estimates of the number of satellite galaxies, we have considered the most conservative reionization scenario from Ref.~\cite{Dooley:2016xkj} to describe $f_{\rm lum}$, with $z_\text{re} = 9.3$, and that different reionization redshifts would correspond to different values of $T_{\rm vir}^{\rm min}$ (if all other parameters are fixed). From this perspective, we can assume to have already approximately marginalized with respect to $T_{\rm vir}^{\rm min}$ (and $\zeta_{\rm UV}$) and thus, these bounds only depend on the IDM cross section. We report the $95\%$~CL exclusion region derived from satellite counts assuming a MW mass of $ M_{\rm MW}^{200} = 0.8 \times 10^{12}~M_{\odot}$ (shaded blue region) and $ M_{\rm MW}^{200} = 2.6 \times 10^{12}~M_{\odot}$ (shaded orange region). These former bounds turn out to be the most stringent ones and help alleviating the strong degeneracy between $\sigma_{\gamma \textrm{DM}}$ and $\zeta_{\rm UV}$ present with $\tau$ and $\bar x_i$ data. Therefore, within the limited number of parameters considered in this work, the preferred values of $\zeta_{\rm UV}$ in IDM scenarios approximately coincide with those in the standard CDM case~\cite{Mesinger:2012ys, Liu:2015txa}.

Recall, however, that the DM model featuring a velocity-independent DM-$\gamma$ scattering cross section considered in our work would correspond to millicharged DM. The stringent limits on this model obtained in Ref.~\cite{Dvorkin:2013cea} translate into the constraint
\begin{align}
\sigma_{ \gamma{\rm DM}} < 8.5 \times 10^{-19} \, \sigma_T \, \left(\frac{\rm GeV}{m_{\rm DM}}\right) ~, 
\end{align}
valid for $m_{\rm DM} \gtrsim$~MeV, and which is several orders of magnitude more constraining than the limit that we have derived from satellite number count.  Now, the effects of DM-photon scattering on the observables considered here are expected to be very similar to those of DM-neutrino interactions. Indeed, the suppression of power at small scales has been shown to be very similar in these two scenarios for the same halo half-mode mass~\cite{Schewtschenko:2014fca}. Lacking dedicated fits for the halo mass function in the DM-neutrino scenario, a rough estimation of the limits on $\sigma_{\nu {\rm DM}}$ can be obtained by rescaling the bounds derived here in terms of the half-mode mass. Noting that very similar half-mode masses are obtained for $\sigma_{\nu {\rm DM}} \sim 1.5 \times \sigma_{\gamma {\rm DM}}$ (see Sec.~\ref{sec:non-standard-dark}), the limits on $\sigma_{\nu {\rm DM}}$ are expected to be in the range $10^{-9}$ to $10^{-10}$ $\times \sigma_T \, (m_{\rm DM}/{\rm GeV})$, which is similar to the constraints from Ly$\alpha$, $\sigma_{\nu {\rm DM}} \lesssim 1.5 \times 10^{-9} \, \sigma_T \, (m_{\rm DM}/{\rm GeV})$~\cite{Wilkinson:2014ksa}.

\subsection{Imprint of IDM and WDM scenarios on the 21~cm signal}
\label{sec:21cm}

Finally, let us briefly discuss the IDM signature in other future cosmological observations. New insights on the EoR ($z = 6 - 12$) and the cosmic dawn ($z \sim 30$) will be provided by the study of the redshifted 21~cm signal (in emission or absorption) from the primordial IGM, associated to the transition between singlet and triplet hyperfine levels of the hydrogen ground state. The observation of the 21~cm line signal represents several advantages compared to traditional cosmic and astrophysical probes (see, e.g., Ref.~\cite{Mao:2008ug, Pritchard:2011xb}). Given that hydrogen is the most abundant element in the Universe, this signal traces the baryonic matter density. By measuring the collective emission from large regions without resolving individual galaxies (intensity mapping), three-dimensional maps of the 21~cm signal could also be obtained. The first generation of radio interferometers testing the 21~cm signal includes the Low Frequency Array (LOFAR)~\cite{vanHaarlem:2013dsa}, the MIT EoR experiment~\cite{Zheng:2014mvp}, the Murchison Widefield Array (MWA)~\cite{Bowman:2012hf, Tingay:2012ps} and the Precision Array for Probing the Epoch of Reionization (PAPER)~\cite{Parsons:2009in} projects, which have already provided upper bounds on the 21~cm signal power spectrum~\cite{Parsons:2009in, Ali:2015uua}. However, these first-generation experiments may not be able to make a definitive detection of the 21~cm signal~\cite{DeBoer:2016tnn}. The next generation of instruments, such as the Hydrogen Epoch of Reionization Array (HERA)~\cite{Pober:2013jna, DeBoer:2016tnn} and the Square Kilometer Array (SKA)~\cite{SKA} will benefit from a larger collecting area and are therefore expected to provide significant measurements of the 21~cm power spectrum. All these experiments will have to deal with foregrounds, that are $\sim 5$ times stronger than the 21~cm cosmological signal (see, e.g., Ref.~\cite{DeBoer:2016tnn} for a summary of different avenues to solve the foregrounds issue). 

\begin{figure}[t]
	\includegraphics[width=0.49\textwidth]{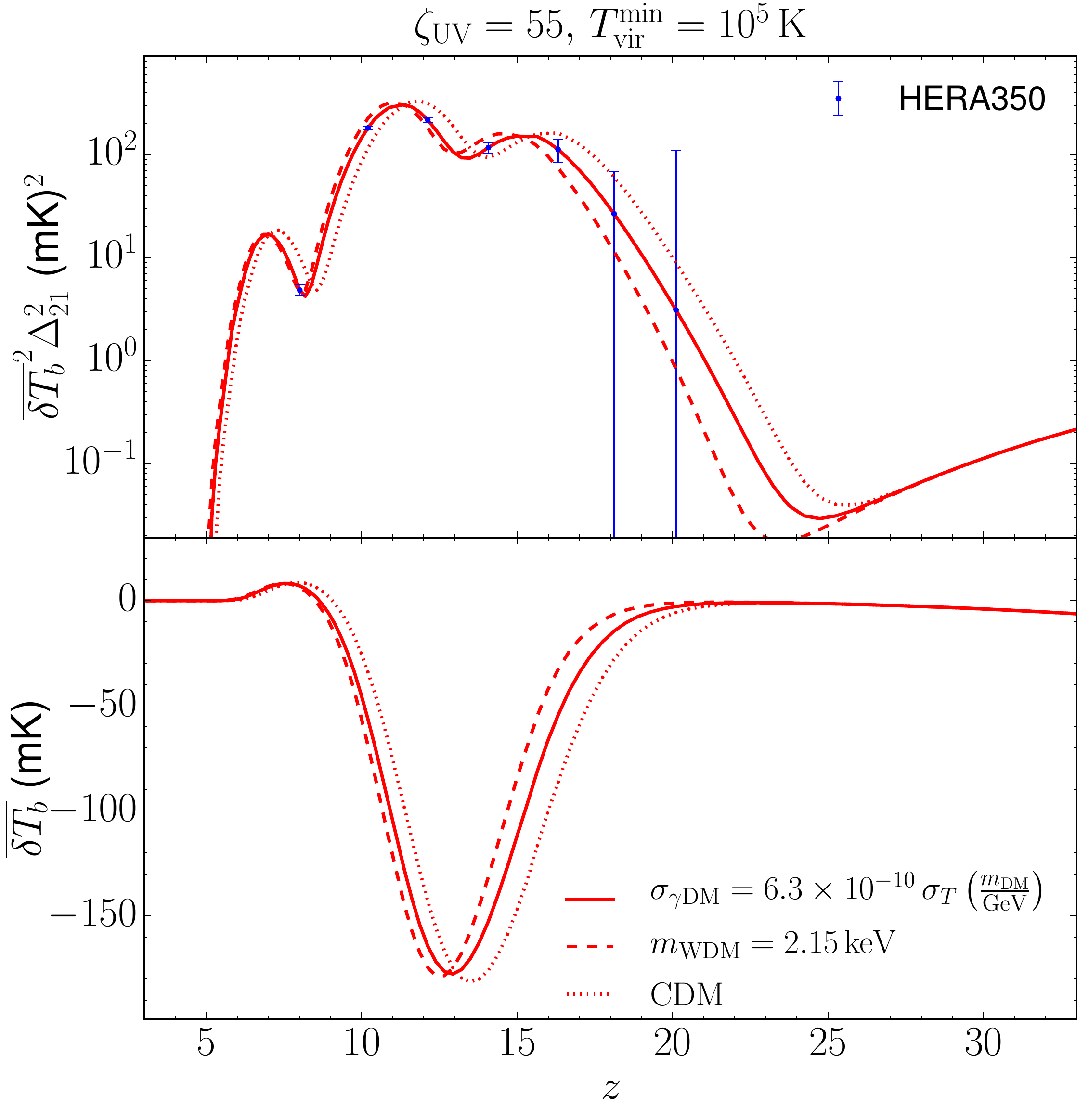} 
	\includegraphics[width=0.49\textwidth]{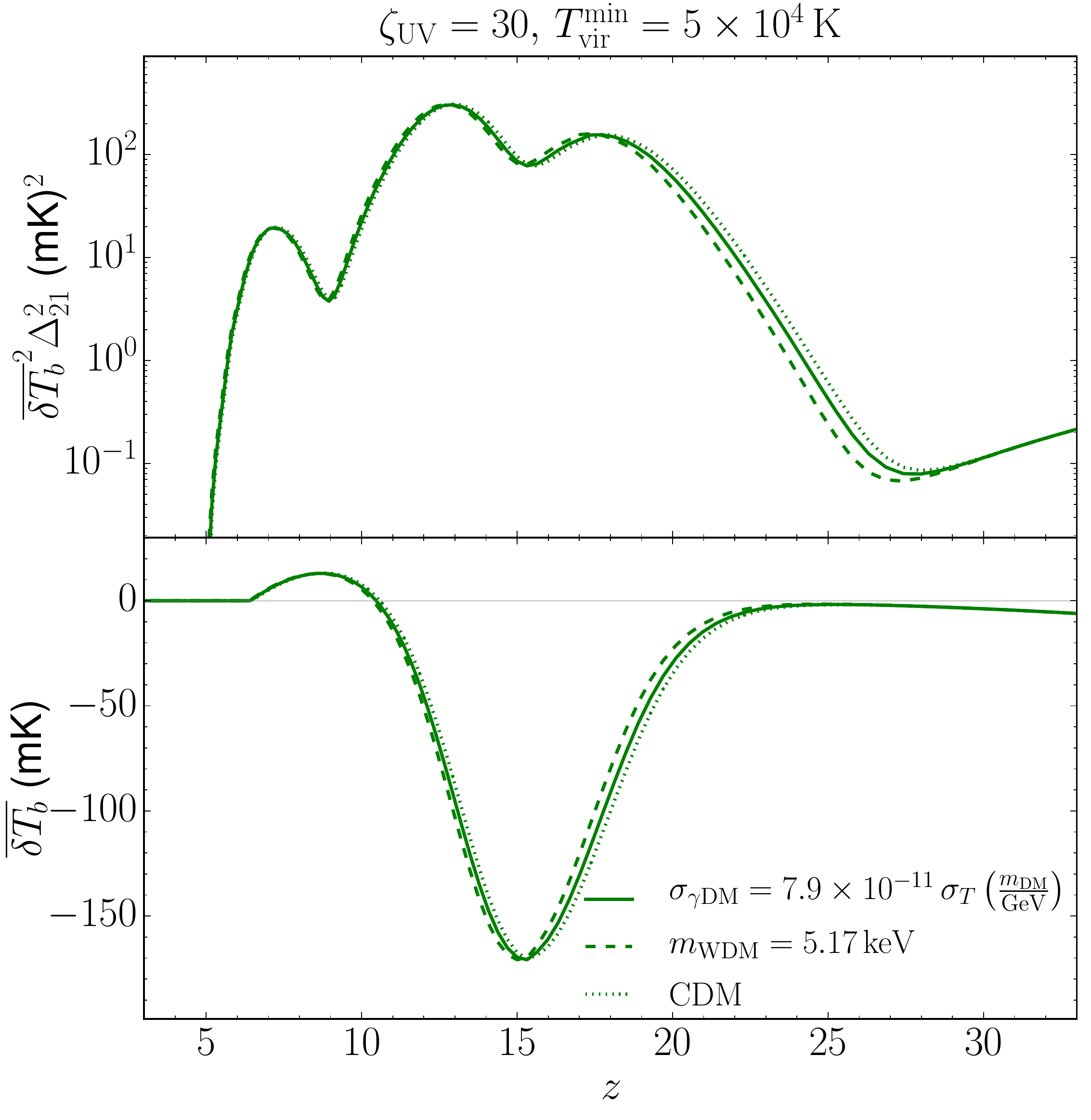}
	\caption{The top panels show the power spectrum of the redshifted 21~cm signal, $\Delta_{21}(z)$, as a function of redshift, $z$, at a fixed scale of $k =0.2$~h/Mpc. The bottom panels depict the associated sky averaged differential brightness temperature, $\delta T_b (z)$. We illustrate the results for CDM (dotted curves), WDM (dashed curves) and IDM (solid curves) scenarios, for the parameter values specified in Tab.~\ref{tab:tabbench}. The IDM and WDM scenarios with a very strong (weak) suppression at small scales are depicted in the left (right) panel. In the left panel we also show the expected sensitivities for the IDM power spectrum from the HERA350 configuration (blue bars) ~\cite{DeBoer:2016tnn}. }
	\label{fig:Tb}
\end{figure}

At this point, let us also mention the recent detection claim of the sky-averaged global 21~cm signal at a redshift $z \sim 18$ by the Experiment to Detect the Global Epoch of Reionization Signatures (EDGES)~\cite{Bowman:2018yin}.  The reported absorption in the measured 21~cm global signal appears to be deeper than that expected in standard CDM scenarios and therefore it cannot be explained by any mechanism that gives rise to heating of the IGM (e.g., DM annihilations~\cite{Evoli:2014pva, Lopez-Honorez:2016sur}). The IDM and WDM scenarios under study in this paper are also unlikely to explain this deep absorption, as they will only shift the absorption dip to later times, leaving unchanged the amplitude of the dip (see lower panels of Fig.~\ref{fig:Tb}). In any case, the EDGES measurement would still need confirmation from observations in other instruments, such as the Large Aperture Experiment to Detect the Dark Ages (LEDA)~\cite{Bernardi:2016pva} or via the detection of the 21~cm power spectrum in the redshift range covered by HERA or SKA.

The brightness of a patch of the IGM relative to the CMB is expressed in terms of the differential brightness temperature, $\delta T_b$, and can be written as~\cite{Madau:1996cs, Furlanetto:2006jb, Pritchard:2011xb, Furlanetto:2015apc}
\begin{align}
\frac{\delta T_b(\nu)}{\textrm{mK} } &\simeq 27  x_\textrm{HI}  (1 + \delta_b) \left( 1 - \frac{T_\textrm{CMB}}{T_{\rm S}}\right) \left( \frac{H}{H+\frac{\partial v_r}{ \partial r}} \right) \, \left( \frac{1+z}{10}\right)^{1/2} \left(\frac{0.15}{\Omega_m h^2} \right)^{1/2} \left( \frac{\Omega_b h^2}{0.023}\right) ~, 
\label{eq:Tbdev}
\end{align}
where $x_\textrm{HI}$ represents the fraction of neutral hydrogen, $\delta_b$ is the baryon overdensity, $\Omega_b h^2$ and $\Omega_m h^2$ refer to the current baryon and matter contribution to the Universe's mass-energy content, $H(z)$ is the Hubble parameter and $\partial v_r / \partial r$ is the comoving gradient of the peculiar velocity along the line of sight. The above expression is exact if $\partial v_r/ \partial r$ is constant over the width of the 21~cm line and $\partial v_r / \partial r\ll H$. The ratio of the populations of the two ground state hyperfine levels of hydrogen is quantified by the spin temperature, $T_{\rm S}$, which is determined by three competing effects~\cite{Hirata:2005mz}: 1) absorption and stimulated emission of CMB photons; 2) atomic collisions, which are important at high redshifts, well before the EoR; and 3) resonant scattering of Ly$\alpha$ photons that turn on with the first sources, the so-called Wouthuysen-Field effect~\cite{Wouthuysen:1952, Field:1958}. The differential brightness temperature power spectrum is defined as
\begin{equation}
\langle  \widetilde{\delta}_{21} (\mathbf{k}, z)  \, \widetilde{ \delta}_{21}^* (\mathbf{k}^\prime, z) \rangle \equiv (2\pi)^3 \, \delta^D (\mathbf{k} - \mathbf{k}^\prime) \, P_{21}(k,z) ~,
\label{eq:P21}
\end{equation}
where $\delta^D$ is the Dirac delta function, the brackets denote an ensemble average, and $\widetilde{\delta}_{21}(\mathbf{k}, z)$ refers to the Fourier transform of ${\delta}_{21}(\mathbf{x}, z)={\delta T}_{b}(\mathbf{x}, z)/ \overline{\delta T_b}(z)-1$ where $\overline{\delta T_b} (z)$ is the sky-averaged differential brightness temperature. The power spectrum $P_{21}(k,z)$ carries information about the correlations in the spin temperature field and is expected to provide the highest signal-to-noise ratio measurement of the 21~cm line around the EoR in the near future. The dimensionless 21~cm differential brightness temperature power spectrum, is defined as
\begin{equation}
\Delta^2_{21} (k,z) = \frac{k^3}{2 \pi^2} \, P_{21}(k,z)~.
\end{equation}

In Fig.~\ref{fig:Tb}, we show the 21~cm power spectrum, $\Delta_{21}^2(k, z)$, multiplied by $\overline{\delta T_b}^2$, as a function of redshift at a fixed scale of $k = 0.2$~h/Mpc (top panels) and the corresponding $\overline{\delta T_b} (z)$ (bottom panels) for CDM (dotted curves) and for the WDM (dashed curves) and IDM (by the solid curves) benchmark scenarios described in Tab.~\ref{tab:tabbench}. These curves have been obtained with the public code {\tt 21cmFast}~\cite{Mesinger:2010ne}. One can see that, for the redshift range shown in Fig.~\ref{fig:Tb} several typical features appear in both $\Delta_{21}^2(z)$ and $\overline{\delta T_b}$. They are directly related to the physical processes driving the 21~cm emission or absorption (see, e.g., Refs.~\cite{Furlanetto:2015apc, Pritchard:2011xb, Mesinger:2013nua}). The power spectra (top panels) show a three-peak structure related, from left to right, to the EoR, the epoch of X-ray heating and the Ly$\alpha$ coupling. The sky-averaged signal appears to be in absorption until X-ray heating processes and in emission until the end of EoR. The expected sensitivities of the HERA350 configuration to 21~cm power spectrum measurements~\cite{DeBoer:2016tnn} are also shown for the IDM power spectrum in one of our benchmark models. These sensitivities have been obtained with the publicly available code {\tt 21cmSense}~\cite{Pober:2013jna, Pober:2012zz}.\footnote{\url{https://github.com/jpober/21cmSense}} We consider in all cases, a total observing time of 1080~hours and a bandwidth of 8~MHz, being both the default parameters in {\tt 21cmSense}. The error bars have been computed for $z = 8, \, 10, \, 12, \, 14, \, 16$ and 20 and include thermal noise plus sample variance. The sensitivity is very good for $z \lesssim 16$, especially around the X-ray heating and the EoR peaks. The left (right) panel of Fig.~\ref{fig:Tb}, corresponds to a WDM/IDM scenario that produces a strong (weak) small scale suppression (see Tab.~\ref{tab:tabbench}), within the allowed region by the measurements discussed in this work. We also show the results for the CDM case, obtained for the very same astrophysical parameters. Overall, WDM and IDM scenarios give rise to a delay in structure formation that shifts to later times the typical features in the 21~cm sky-averaged signal and power spectrum. The same effect was observed when we discussed reionization in Section~\ref{sec:reio-cons} and it is more pronounced for the WDM case.

For a fixed value of the half-mode mass, there is a fall of power in $\Delta_{21}^2$ for IDM and WDM models, and a corresponding vanishing $\overline{\delta T_b}$, at very similar redshift, $z \sim 8$, driven by the end of reionization, as expected from Fig.~\ref{fig:ioWDMIDM}. However, comparing IDM and WDM scenarios, the difference in the shifts of the second (X-ray heating) and third (Ly$\alpha$ coupling) peaks is more pronounced than in the first peak (EoR). Therefore, the differences between IDM and WDM cannot be simply compensated by a shift in the redshift evolution of the 21~cm signal (as it is the case, for instance, when changing $T_{\rm vir}^{\rm min}$~\cite{Lopez-Honorez:2016sur}). The larger number of low mass halos in the IDM case shifts the different milestone epochs of the 21~cm signal in a non-trivial and different way from the WDM case. This is very interesting, as points to a distinctive signature to disentangle these two scenarios, otherwise difficult to distinguish. Thus, in principle, barring out of the discussion astrophysical uncertainties, one could potentially be able to differentiate between IDM and WDM scenarios by studying the redshift interval between the reionization and the X-ray heating processes, although a dedicated analysis is beyond the scope of this paper. Obviously, for very small (large) values of the IDM cross section (WDM mass), distinguishing any of these scenarios (WDM, IDM and CDM) from each other would be even more challenging (see the right panels of Fig.~\ref{fig:Tb}).

\section{Summary and conclusions}
\label{sec:conclusions}

A number of observational probes of our Universe at galactic and subgalactic scales may require a modification to the standard CDM paradigm. These small-scale measurements indicate that \textit{(a)} dwarf galaxies are hosted by halos that are less massive than those predicted in the CDM numerical simulations, and \textit{(b)} the observed number of satellite galaxies that orbit close to the MW is smaller than that predicted within standard CDM cosmology. Possible avenues to overcome these problems have been proposed in the literature~\cite{Wang:2016rio, Lovell:2016nkp}, ranging from lowering the total mass of the satellites by baryon or supernovae feedback effects~\cite{Sawala:2012cn, Sawala:2015cdf, Fattahi:2016nld}, by changes in the numerical simulations~\cite{Polisensky:2013ppa} or by modifying the DM model sector. Focusing on this last solution, possible modifications to the standard CDM paradigm include IDM scenarios (see, e.g., Ref.~\cite{Schewtschenko:2015rno}) and WDM candidates (such as sterile neutrinos)~\cite{Lovell:2011rd, Lovell:2013ola, Lovell:2015psz}, which attracted recently more attention due to possible hints in X-ray data~\cite{Bulbul:2014sua, Boyarsky:2014jta, Boyarsky:2014ska, Cappelluti:2017ywp}. Here, we have focused on IDM scenarios (and its comparison with WDM scenarios), as they provide a possible solution to the small-scale crisis via collisional damping effects that would suppress the amount of small-scale structures. Nevertheless, the suppression of small structures would also directly impact on different cosmological observables. In this work we consider the impact of these non-CDM scenarios on reionization-related observables, along with constraints from the number of observed satellite galaxies of the MW. 

After describing the halo mass function obtained from numerical simulations in the case of IDM (and WDM) scenarios (Section~\ref{sec:non-standard-dark}), we first study the effects on the ionization history of the Universe within this kind of scenarios (Fig.~\ref{fig:ioWDMIDM}), parameterized by the ratio of the DM-photon elastic scattering cross section, $\sigma_{\gamma {\rm DM}}$: the IDM collisional damping would wash out small-scale overdensities, delaying the onset of reionization and thus, would affect reionization-related observables. We have considered three different types of measurements in order to constrain the IDM scenario. Namely, for our analyses we include the CMB Planck integrated optical depth $\tau$, the Gunn-Peterson optical depth low-$z$ data and high-$z$ constraints from Ly$\alpha$ emission (Section~\ref{sec:reio}). Nevertheless, the effects of collisional damping in IDM scenarios are somewhat degenerate with a number of (uncertain) astrophysical parameters governing the ionization processes in the Universe, as the minimum virial temperature $T_{\rm vir}^{\rm min}$ and the UV ionization efficiency, $\zeta_{\rm UV}$.  Indeed, we have explicitly shown the degeneracy between $\zeta_{\rm UV}$ and $\sigma_{\gamma {\rm DM}}$ (Fig.~\ref{fig:Chi2xe}). Although current data are not precise enough to disentangle it, the combination of future low-$z$ Gunn-Peterson and high-$z$ measurements of the ionization history provides a promising tool to discard IDM scenarios. All in all, at present, the combination of reionization observables allows us to set an upper bound on IDM cross section,  $\sigma_{\gamma \rm{DM}} < 4 \times 10^{-9} \, \sigma_T  \times \left( {m_{\rm DM}}/{\rm GeV}\right)$ at $95\%$~CL. 

These constraints are complemented by studying the predicted number of MW satellites as a function of the IDM cross section, which are also compared with the results in WDM scenarios using the DM mass as the free parameter (Section~\ref{sec:Nsat}). We profit here from the recent updates in the statistics of MW satellite galaxy number counts: the eleven classical MW satellites, the seventeen objects discovered by DES~\cite{Bechtol:2015cbp, Drlica-Wagner:2015ufc}, the seventeen satellites detected by SDSS~\cite{Ahn:2012fh, Koposov:2009ru} and the nine objects found in other surveys~\cite{Newton:2017xqg}. Using the latest estimation for the number of satellite galaxies, $N_\text{gal}$, of the MW, which accounts for the latest discoveries and recent N-body simulations, a lower bound was obtained, $N_\text{gal} > 85$ at 95\%~ CL~\cite{Newton:2017xqg}. This result is confronted with the predictions for IDM (and WDM) scenarios in order to constrain the IDM cross section (and WDM mass). For these estimates, we follow the analytical approach of Ref.~\cite{Kim:2017iwr} and take into account the probability of a subhalo to actually host a galaxy (Section~\ref{sec:Nsat}).  Our most conservative 95\%~CL upper limit on the IDM cross section is found for the highest value of the MW mass we consider ($M_{\rm MW}^{200} = 2.6 \times 10^{12}~ M_{\odot}$), $\sigma_{\gamma \rm{DM}} < 8 \times 10^{-10} \, \sigma_T \, \left(m_{\rm DM}/{\rm GeV}\right)$ (Fig.~\ref{fig:Sat_res}), which implies an order of magnitude improvement over the limits on the DM-photon elastic scattering cross section obtained in Ref.~\cite{Boehm:2014vja}. Nevertheless, we have also shown that under the same assumptions and using the same set of data, we can approximately recover the results of this previous work (Fig.~\ref{fig:comp_IDM}). In case that the MW mass is $ M_{\rm MW}^{200} = 0.8 \times 10^{12}~M_{\odot}$, the resulting 95\%~CL upper limit on the IDM cross section is found to be $\sigma_{\gamma \rm{DM}} < 9 \times 10^{-11} \, \sigma_T \, \left(m_{\rm DM}/{\rm GeV}\right)$. Note that recent analyses using Gaia data point toward the lower side range of MW masses that we consider~\cite{2018arXiv180409381G,2018arXiv180501408P,2018arXiv180500908F,2018arXiv180411348W}.  

Throughout this paper, we have also discussed the impact of another non-CDM scenario, as it is the case of WDM candidates. This scenario could also potentially solve discrepancies observed at galactic and subgalactic scales. Due to their free-streaming, WDM particles would also lead to a suppression in the small-scale power spectrum. Indeed, it is possible to establish an approximate connection between the power spectrum in these two non-CDM scenarios (Fig.~\ref{fig:transfer}), in terms of the scale at which the transfer function is reduced by half, although the connection between the number of low-mass halos requires additional corrections (Section~\ref{sec:non-standard-dark}). Therefore, in a similar way we can obtain an upper limit on the IDM cross section, we can also obtain a lower limit on the WDM mass. Using MW satellite galaxy counts and similarly as done for IDM, we get $m_{\rm WDM} > 2.8$~keV at 95\%~CL for a thermal candidate (or $m_s > 16$~keV at 95\%~CL for a non-resonantly produced sterile neutrino) for $M_{\rm MW}^{200} = 2.6 \times 10^{12}~ M_{\odot}$ (Fig.~\ref{fig:Sat_res}).

These results indicate that currently, among the different data sets used in this analysis, the most restricting one for the non-CDM scenarios considered is the estimated number of MW satellite galaxies (Fig.~\ref{fig:Chi2xet}). Notice, though, that since DM-photon interactions of the type considered in this work imply DM-nucleon interactions with a Rutherford-like cross section, Ly$\alpha$ and CMB data set more stringent constraints on the latter than the bounds obtained here. We argue, however, that the method use could be applied to any other IDM model with DM-radiation interaction. In particular, in the case of DM-neutrino interactions, the bounds derived in this work are among the most constraining limits on  such scenario.

Note that the approximate connection that can be established between IDM and WDM scenarios also implies that they give rise to very similar effects in the observables we have considered and thus, the IDM cross section is (almost) fully degenerate with the WDM mass. Therefore, distinguishing them using this type of data would be a very challenging task. Nevertheless, we have also briefly discussed the potential distinctive signatures of these scenarios in future cosmological measurements of the 21~cm hydrogen line (Section~\ref{sec:21cm}), which will probe the Universe evolution beyond the EoR. In particular, IDM scenarios could give rise to a larger relative shift in redshift between the characteristic features of the 21~cm power spectrum and of the sky-averaged global brightness temperature than WDM scenarios with the same half-mode mass. Therefore, future, very large radio interferometers, as HERA or SKA, may also be important tools to shed light on the nature of DM particles and on their precise clustering and interacting properties.

\section*{Acknowledgments}
We would like to thank J.~Pober for providing us with the HERA350 configuration files, C. B{\oe}hm for discussions and P.~Champion for useful cross-checks.  ME is supported by Spanish Grant FPU13/03111 of MECD. LLH is supported by the Fonds National de la Recherche Scientifique, by the Belgian Federal Science Policy Office through the Interuniversity Attraction Pole P7/37 and by the Vrije Universiteit Brussel through the Strategic Research Program \textit{High-Energy Physics}. ME, OM, PVD are supported by PROMETEO II/2014/050 and by the Spanish Grants FPA2014--57816-P and FPA2017-85985-P of the MINECO.  SPR is supported by a Ram\'on y Cajal contract, by the Spanish MINECO under grants FPA2017-84543-P and FPA2014-54459-P and by the Generalitat Valenciana under grant PROMETEOII/2014/049. ME, OM, SPR and PVD are also supported by the MINECO Grant SEV-2014-0398 and by the European Union's Horizon 2020 research and innovation program under the Marie Sk\l odowska-Curie grant agreements No. 690575 and 674896.  SPR is also partially supported by the Portuguese FCT through the CFTP-FCT Unit 777 (PEst-OE/FIS/UI0777/2013).

\bibliographystyle{JHEP}
\bibliography{biblio}

\end{document}